\documentclass[twocolumn,showpacs,preprintnumbers,amsmath,amssymb,superscriptaddress, prc]{revtex4-1}

\usepackage{rotating}
\usepackage{graphicx}
\usepackage{dcolumn}
\usepackage{bm}
\usepackage{mathtools}
\usepackage{amsmath}
\usepackage{textcomp}
\usepackage{longtable}
\usepackage{color}
\usepackage{hyperref}
\hypersetup{
 colorlinks,
 linkcolor={blue},
 citecolor={blue},
 urlcolor={blue}
}

\begin{document}

\title{Effects of one valence proton on seniority and angular momentum of neutrons in neutron-rich $^{122-131}_{51}$Sb isotopes
}

\newcommand{\IPHC}{Universit\'e de Strasbourg, CNRS, IPHC UMR 7178, F-67000 Strasbourg, France}
\newcommand{\IPNL}{Universit\'e de Lyon, Universit\'e Lyon-1, CNRS/IN2P3, UMR5822, IPNL, 4 Rue Enrico Fermi, F-69622 Villeurbanne Cedex, France}
\newcommand{\GANIL}{GANIL, CEA/DRF-CNRS/IN2P3, Bd Henri Becquerel, BP 55027, F-14076 Caen Cedex 05, France}
\newcommand{\CSNSM}{CSNSM, Univ. Paris-Sud, CNRS/IN2P3, Universit\'e Paris-Saclay, 91405 Orsay, France}
\newcommand{\TUDarmstadt}{Institut f\"ur Kernphysik, Technische Universit\"at Darmstadt, D-64289 Darmstadt, Germany}
\newcommand{\IFIC}{Instituto de F\'isica Corpuscular, CSIC-Universitat de Val\`encia, E-46980 Val\`encia, Spain}
\newcommand{\Canada}{Department of Chemistry, Simon Fraser University, Burnaby, British Columbia, Canada}
\newcommand{\Legnaro}{INFN, Laboratori Nazionali di Legnaro, Via Romea 4, I-35020 Legnaro, Italy}
\newcommand{\IPN}{Institut de Physique Nucl\'eaire, IN2P3-CNRS, Univ. Paris Sud, Universit\'e Paris Saclay, 91406 Orsay Cedex, France}
\newcommand{\Debrecen}{Institute for Nuclear Research of the Hungarian Academy of Sciences, Pf.51, H-4001, Debrecen, Hungary}
\newcommand{\Somerset}{iThemba LABS, National Research Foundation, P.O.Box 722, Somerset West,7129 South Africa}
\newcommand{\Padova}{INFN Sezione di Padova, I-35131 Padova, Italy}
\newcommand{\UPadova}{Dipartimento di Fisica e Astronomia dell'Universit\`a di Padova, I-35131 Padova, Italy}
\newcommand{\GSI}{GSI, Helmholtzzentrum f\"ur Schwerionenforschung GmbH, D-64291 Darmstadt, Germany}
\newcommand{\Milano}{INFN, Sezione di Milano, I-20133 Milano, Italy}
\newcommand{\LPSC}{LPSC, Universit\'e Grenoble-Alpes, CNRS/IN2P3, 38026 Grenoble Cedex, France}
\newcommand{\IRFU}{IRFU, CEA/DRF, Centre CEA de Saclay, F-91191 Gif-sur-Yvette Cedex, France}
\newcommand{\UMilano}{Dipartimento di Fisica, Universit\`a di Milano, I-20133 Milano, Italy}
\newcommand{\STFC}{STFC Daresbury Laboratory, Daresbury, Warrington, WA4 4AD, UK}
\newcommand{\LBNL}{Nuclear Science Division, Lawrence Berkeley National Laboratory, Berkeley, California 94720, USA}
\newcommand{\IFJ}{Institute of Nuclear Physics PAN, 31-342 Krak\'ow, Poland}
\newcommand{\VECC}{Variable Energy Cyclotron Centre, 1/AF Bidhan Nagar, Kolkata 700064, India}
\newcommand{\TIFR}{Department of Nuclear and Atomic Physics, Tata Institute of Fundamental Research,Mumbai, 400005, India}
\newcommand{\ILL}{Institut Laue-Langevin, F-38042 Grenoble Cedex, France}
\newcommand{\HBNI}{Homi Bhabha National Institute, Training School Complex, Anushaktinagar, Mumbai-400094, India}

\author{S. Biswas}
\email{biswas@ganil.fr}
\affiliation{\GANIL}

\author{A.~Lemasson}
\email{lemasson@ganil.fr}
\affiliation{\GANIL}

\author{M.~Rejmund}
\affiliation{\GANIL}

\author{A.~Navin}
\affiliation{\GANIL}

\author{Y.H.~Kim}
\altaffiliation[Present address: ]{\ILL}
\affiliation{\GANIL}

\author{C.~Michelagnoli}
\altaffiliation[Present address: ]{\ILL}
\affiliation{\GANIL}

\author{I.~Stefan}
\affiliation{\IPN}

\author{R.~Banik}
\affiliation{\VECC}
\affiliation{\HBNI}

\author{P. Bednarczyk}
\affiliation{\IFJ}

\author{S.~Bhattacharya}
\affiliation{\VECC}
\affiliation{\HBNI}

\author{S.~Bhattacharyya}
\affiliation{\VECC}
\affiliation{\HBNI}

\author{E.~Cl\'{e}ment}
\affiliation{\GANIL}

\author{H. L. Crawford}
\affiliation{\LBNL}

\author{G.~de~France}
\affiliation{\GANIL}

\author{P. Fallon}
\affiliation{\LBNL}

\author{G.~Fr\'{e}mont}
\affiliation{\GANIL}

\author{J.~Goupil}
\affiliation{\GANIL}

\author{B.~Jacquot}
\affiliation{\GANIL}

\author{H.J.~Li}
\affiliation{\GANIL}

\author{J.~Ljungvall}
\affiliation{\CSNSM}


\author{A. Maj}
\affiliation{\IFJ}

\author{L.~M\'enager}
\affiliation{\GANIL}

\author{V.~Morel}
\affiliation{\GANIL}

\author{R.~Palit}
\affiliation{\TIFR}

\author{R.M.~P\'erez-Vidal}
\affiliation{\IFIC}

\author{J.~Ropert}
\affiliation{\GANIL}


%
%
%
%
%
%
\author{D.~Barrientos}
\affiliation{CERN, CH-1211 Geneva 23 (Switzerland)}
%
%
%
%
%
%
%
%
%
%
%
%
%
%
\author{G.~Benzoni}
\affiliation{\Milano}
\author{B.~Birkenbach}
\affiliation{Institut f\"ur Kernphysik, Universit\"at zu K\"oln, Z\"ulpicher Str. 77, D-50937 K\"oln, Germany}
%
%
%
%
\author{A.J.~Boston}
\affiliation{Oliver Lodge Laboratory, The University of Liverpool, Liverpool, L69 7ZE, UK}
\author{H.C.~Boston}
\affiliation{Oliver Lodge Laboratory, The University of Liverpool, Liverpool, L69 7ZE, UK}
%
%
%
%
%
%
%
%
%
%
%
%
\author{B.~Cederwall}
\affiliation{Department of Physics, Royal Institute of Technology, SE-10691 Stockholm, Sweden}
%
%
%
%
%
%
%
%
\author{J.~Collado}
\affiliation{Departamento de Ingenier\'ia Electr\'onica, Universitat de Valencia, Burjassot, Valencia, Spain}
%
%
%
%
%
\author{D.M.~Cullen}
\affiliation{Nuclear Physics Group, Schuster Laboratory, University of Manchester, Manchester, M13 9PL, UK}
\author{P.~D\'esesquelles}
\affiliation{\CSNSM}
%
%
%
%
%
%
%
%
\author{C.~Domingo-Pardo}
\affiliation{\IFIC}
%
%
%
%
%
%
\author{J.~Dudouet}
\affiliation{\CSNSM}
\affiliation{\IPNL}
\author{J.~Eberth}
\affiliation{Institut f\"ur Kernphysik, Universit\"at zu K\"oln, Z\"ulpicher Str. 77, D-50937 K\"oln, Germany}
%
%
%
%
%
%
%
%
%
%
%
%
%
%
\author{V.~Gonz\'alez}
\affiliation{Departamento de Ingenier\'ia Electr\'onica, Universitat de Valencia, Burjassot, Valencia, Spain}
%
%
%
%
%
%
%
%
%
\author{L.~J.~Harkness-Brennan}
\affiliation{Oliver Lodge Laboratory, The University of Liverpool, Liverpool, L69 7ZE, UK}
\author{H.~Hess}
\affiliation{Institut f\"ur Kernphysik, Universit\"at zu K\"oln, Z\"ulpicher Str. 77, D-50937 K\"oln, Germany}
%
%
%
%
%
%
%
%
%
%
%
%
%
%
\author{A.~Jungclaus}
\affiliation{Instituto de Estructura de la Materia, CSIC, Madrid, E-28006 Madrid, Spain}
%
%
%
%
%
%
%
%
%
%
%
%
%
%
\author{W.~Korten}
\affiliation{\IRFU}
\author{M.~Labiche}
\affiliation{\STFC}
%
%
%
%
%
%
\author{A.~Lefevre}
\affiliation{\GANIL}
%
%
%
%
%
%
%
%
%
%
%
%
%
%
%
%
%
%
%
%
%
%
%
%
%
%
%
%
%
\author{R.~Menegazzo}
\affiliation{\Padova}
\author{D.~Mengoni}
\affiliation{\Padova}
\affiliation{\UPadova}
%
%
\author{B.~Million}
\affiliation{\Milano}
%
%
%
%
%
%
%
%
%
%
\author{D.~R.~Napoli}
\affiliation{\Legnaro}
%
%
%
%
%
%
%
%
%
%
%
%
%
%
\author{A.~Pullia}
\affiliation{\Milano}
\affiliation{\UMilano}
\author{B.~Quintana}
\affiliation{Laboratorio de Radiaciones Ionizantes, Universidad de Salamanca, E-37008 Salamanca, Spain}
%
%
%
%
%
\author{D.~Ralet}
\affiliation{\CSNSM}
\affiliation{\TUDarmstadt}
\affiliation{\GSI}
\author{F.~Recchia}
\affiliation{\Padova}
\affiliation{\UPadova}
%
%
%
%
%
\author{P.~Reiter}
\affiliation{Institut f\"ur Kernphysik, Universit\"at zu K\"oln, Z\"ulpicher Str. 77, D-50937 K\"oln, Germany}
%
%
%
%
%
%
\author{F.~Saillant}
\affiliation{\GANIL}
\author{M.D.~Salsac}
\affiliation{\IRFU}
\author{E.~Sanchis}
\affiliation{Departamento de Ingenier\'ia Electr\'onica, Universitat de Valencia, Burjassot, Valencia, Spain}
%
%
%
%
%
%
%
%
%
%
%
%
\author{O.~Stezowski}
\affiliation{\IPNL}
\author{Ch.~Theisen}
\affiliation{\IRFU}
%
%
%
%
%
%
%
%
\author{J.J.~Valiente-Dob\'on}
\affiliation{\Legnaro}
%
%
%
%
%
%
%
%
%
%
%
%
\author{M.~Zieli\'nska}
\affiliation{\IRFU}

\date{\today}

\begin{abstract}

{\bf Background:} Levels fulfilling seniority scheme and relevant isomers are commonly observed features in semi-magic nuclei, for example in Sn isotopes ($Z = 50$). Seniority isomers in Sn, with dominantly pure neutron configurations, directly probe the underlying neutron-neutron ($\nu\nu$) interaction. Further, an addition of a valence proton particle or hole, through neutron-proton ($\nu\pi$) interaction, affects the neutron seniority as well as the angular momentum. 

{\bf Purpose:} Benchmark the reproducibility of the experimental observables, like the excitation energies ($E_{X}$) and the reduced electric quadrupole transition probabilities ($B(E2)$), with the results obtained from shell model interactions for neutron-rich Sn and Sb isotopes with $N <$ 82. Study the sensitivity of the aforementioned experimental observables to the model interaction components. Further, explore from a microscopic point of view the structural similarity between the isomers in Sn and Sb, and thus the importance of the valence proton.

{\bf Methods:} The neutron-rich $^{122-131}$Sb isotopes were produced as fission fragments in the reaction $^{9}$Be($^{238}$U,~f) with 6.2 MeV/u beam energy. An unique setup, consisting of AGATA, VAMOS++ and EXOGAM detectors, was used which enabled the prompt-delayed gamma-ray ($\gamma$) spectroscopy of fission fragments in the time range of 100 ns - 200 $\mu$s.

{\bf Results:} New isomers, prompt and delayed transitions were established in the even-A $^{122-130}$Sb isotopes. In the odd-A $^{123-131}$Sb isotopes, new prompt and delayed $\gamma$-ray transitions were identified, in addition to the confirmation of the previously known isomers.  The half-lives of the isomeric states and the $B(E2)$ transition probabilities of the observed transitions depopulating these isomers were extracted.

{\bf Conclusions:} The experimental data was compared with the theoretical results obtained in the framework of Large-Scale Shell-Model (LSSM) calculations in a restricted model space. Modifications of several components of the shell model interaction were introduced to obtain a consistent agreement with the excitation energies and the $B(E2)$ transition probabilities in neutron-rich Sn and Sb isotopes. The isomeric configurations in Sn and Sb were found to be relatively pure. Further, the calculations revealed that the presence of a single valence proton, mainly in the $g_{7/2}$ orbital in Sb isotopes, leads to significant mixing (due to the $\nu\pi$ interaction) of: (i) the neutron seniorities ($\upsilon_{\nu}$) and (ii) the neutron angular momentum ($I_{\nu}$). The above features have a weak impact on the excitation energies, but have an important impact on the $B(E2)$ transition probabilities. In addition, a constancy of the relative excitation energies irrespective of neutron seniority and neutron number in Sn and Sb was observed. 

\end{abstract}

\maketitle

 
\section{\label{sec:Intro}Introduction}

The Sn isotopes, with basically a spherical shape, span between two doubly magic nuclei, the neutron-deficient $^{100}$Sn ($Z~=~N~=~50$) and the neutron-rich $^{132}$Sn ($Z~=~50$,~$N~=~82$). A constant trend, as a function of neutron number, in the excitation energies ($E_{X}$) of the $2^{+}$ state and the parabolic shape (with a dip around $N$~=~66) of the reduced electric quadrupole transition probabilities ($B(E2; 2^{+} \rightarrow 0^{+})$), was observed for the even-A Sn isotopes~\cite{casten,nudat}. The observed trend has recently been clarified using state-of-the-art Monte-Carlo Shell-Model (MCSM) calculations with a large model space, where the dip in the $B(E2)$ was shown to correspond to a novel shape evolution from moderately deformed phase to pairing (seniority) phase around $N = 66$~\cite{to18}. These calculations required huge computational resources, hence only $0^{+}$, $2^{+}$, and $4^{+}$ states were calculated for fitting purpose. Recently, the parabolic behaviour of the $B(E2; 10^{+} \rightarrow 8^{+})$ in even-A $^{116-130}$Sn was described in the generalised seniority ($\upsilon$) scheme in Ref.~\cite{ma16}, showing the necessity of the incorporation of configuration mixing of neutron $\nu h_{11/2}$, $\nu d_{3/2}$ and $\nu s_{1/2}$ orbitals instead of a pure $\nu h_{11/2}$ orbital, with generalised seniority being pure $\upsilon_{\nu}$ = 2 ($\upsilon_{\nu}$ stands for neutron seniority, which refers to the number of unpaired neutrons). It is known that seniority is a good quantum number for $l_{j} \le f_{7/2}$. This becomes a partial symmetry in the case of $l_{j} = g_{9/2}$ in $^{72,74}$Ni~\cite{mo18}. In these isotopes, the ($6^{+}$)$_{\upsilon_{\nu} = 4}$ state has a lower excitation energy than the ($6^{+}$)$_{\upsilon_{\nu} = 2}$, leading to a disappearance of the seniority ($8^{+}$)$_{\upsilon_{\nu} = 2}$ isomers. At present, there is no indication of disappearance of ($10^{+}$)$_{\upsilon_{\nu} = 2}$ seniority isomers in the Sn isotopes ($l_{j} = h_{11/2}$), indicating that the $8^{+}$ state in Sn is still a pure $\upsilon_{\nu}$ = 2 state. Recent results on the low-lying high-spin states in even-A $^{124-128}$Sb ($Z = 51$) and $^{122-126}$In ($Z = 49$) suggested that there is a significant mixing of the seniorities, $\upsilon_{\nu}$ = 1 and 3, due to the presence of a single valence proton particle/hole in the $g_{7/2}$/$g_{9/2}$ orbital~\cite{re16}. It was shown in Ref.~\cite{re16} that the interaction of a single proton particle/hole in the $g_{7/2}$/$g_{9/2}$ spin-orbit partners in Sb/In isotopes, respectively with the neutrons energetically favoured the state with one broken pair of neutrons (breaking of the seniority) over a state with no broken pairs for the low-lying high-spin states. 

The experimental study of nuclei near the shell closures serves as an important test bench for directly probing the nucleon-nucleon interaction. This acts as important input for existing shell-model interactions and allows to improve the predictive power of such calculations. In particular, the mass A $\sim$ 130 region, near shell closure, is an interesting area of research in contemporary nuclear physics, as $^{132}$Sn is so far the heaviest neutron-rich unstable doubly magic nucleus and it is possible to access this region with the recent advent of new accelerator facilities and large detector arrays~\cite{bh01, co10, jo10, ro18}. The excitation energies ($E_{X}$) and the transition strengths ($B(E2)/B(M1)$) of neighbouring nuclei near $^{132}$Sn act as a testing ground to shell-model calculations and the associated interactions~\cite{co15, te15, bi16, re16, wa17, ki18}.

Even-A $^{118-130}$Sn isotopes possess $7^{-}$ and $10^{+}$ isomers, with dominant neutron $\nu h_{11/2}^{-1}d_{3/2}^{-1}$ and $\nu h_{11/2}^{-2}$ configurations, respectively~\cite{fo81, pi11, as12, is14}. Similarly, odd-A $^{119-129}$Sn isotopes display $23/2^{+}$ and $27/2^{-}$ isomers with dominant $\nu h_{11/2}^{-2}d_{3/2}^{-1}$ and $\nu h_{11/2}^{-3}$ configurations, respectively~\cite{lo08, as12, is16}, an additional hole in the $\nu h_{11/2}$ orbital coupled to the counterparts, $7^{-}$ and $10^{+}$ isomers in even-A Sn, respectively. The experimental data from these nuclei, due to proton shell closure, probe the neutron-neutron ($\nu\nu$) interaction. Odd-A $^{121-131}$Sb isotopes have $19/2^{-}$ and $23/2^{+}$ isomers with dominant $\pi g_{7/2} \nu h_{11/2}^{-1}d_{3/2}^{-1}$ and $\pi g_{7/2} \nu h_{11/2}^{-2}$ configurations ~\cite{ju07, wa09, wa09epja, ge03, ge00}, an additional proton particle in $g_{7/2}$ coupled to the $7^{-}$ and $10^{+}$ isomers in even-A Sn, respectively. The isomers in even-A $^{122-128}$Sb isotopes, except for $^{130}$Sb, were not known and are newly observed in this work. The experimental data on Sb isotopes, in addition to the information obtained for Sn, probe the neutron-proton ($\nu\pi$) interaction. The present manuscript discusses the four different cases of configurational similarities between Sn and Sb:\\
{\bf Case I}: $10^{+} \rightarrow 8^{+}$ ($\nu h_{11/2}^{-2}$) in even-A Sn  and  $23/2^{+} \rightarrow 19/2^{+}$ ($\pi g_{7/2} \nu h_{11/2}^{-2}$) in odd-A Sb \\
{\bf Case II}: $7^{-} \rightarrow 5^{-}$ ($\nu h_{11/2}^{-1}d_{3/2}^{-1}$) in even-A Sn  and  $19/2^{-} \rightarrow 15/2^{-}$ ($\pi g_{7/2} \nu h_{11/2}^{-1}d_{3/2}^{-1}$) in odd-A Sb \\
{\bf Case III}: $23/2^{+} \rightarrow 19/2^{+}$ ($\nu h_{11/2}^{-2}d_{3/2}^{-1}$) in odd-A Sn  and  $13^{+} \rightarrow 11^{+}$ ($\pi g_{7/2} \nu h_{11/2}^{-2}d_{3/2}^{-1}$) in even-A Sb \\
{\bf Case IV}: $27/2^{-} \rightarrow 23/2^{-}$ ($\nu h_{11/2}^{-3}$) in odd-A Sn and $16^{-} \rightarrow 14^{-}$ ($\pi g_{7/2}\nu h_{11/2}^{-3}$) in even-A Sb.

\begin{turnpage}
\begin{figure*}[]
\vspace{-10cm}
\includegraphics[width=2.8\columnwidth]{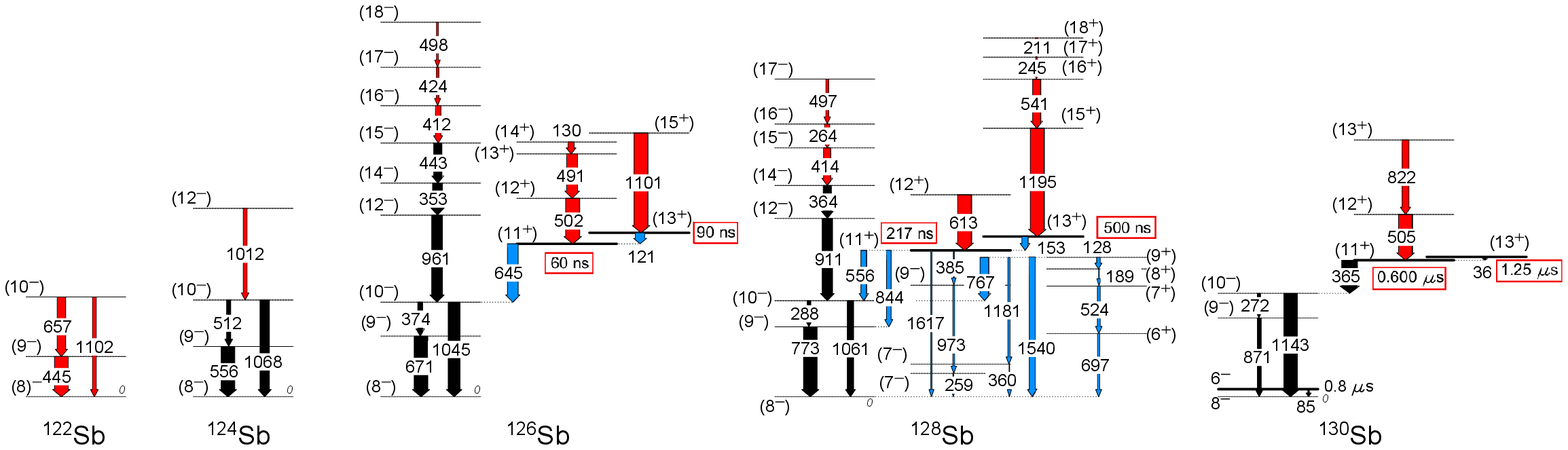}
\includegraphics[width=2.8\columnwidth]{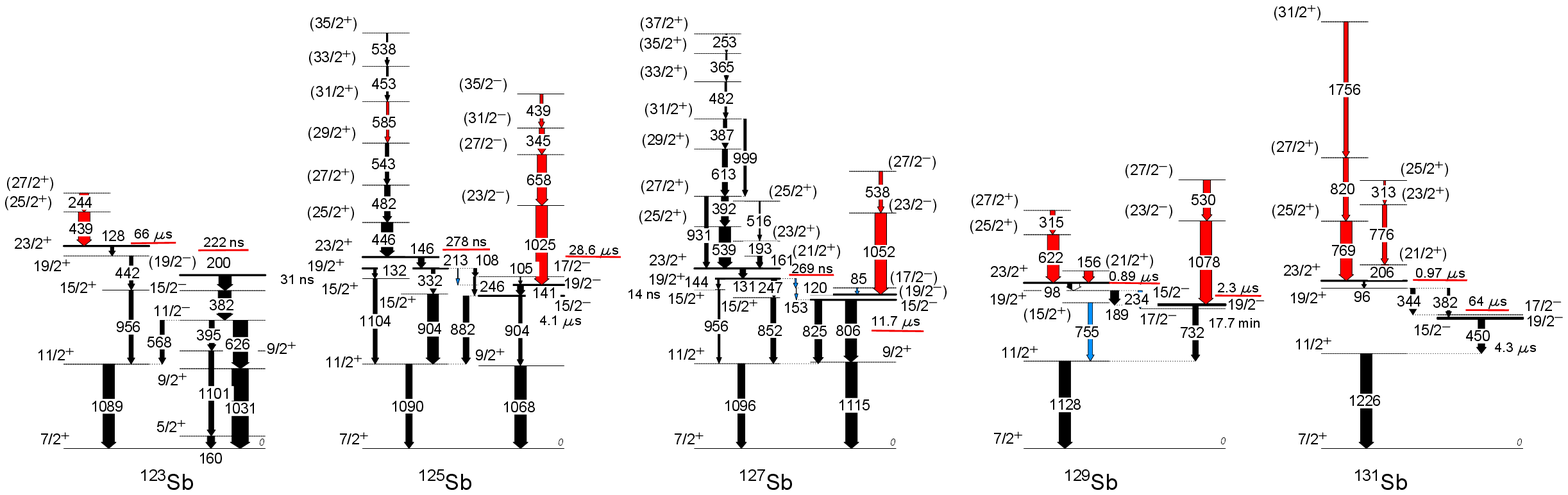}
\caption{\label{fig:sb_fig1} (Color online) The level schemes of $^{122-131}$Sb. The newly observed $\gamma$-ray transitions above and below the isomer are indicated in red and blue, respectively. The width of the arrows represent the intensity of the transitions.  The isomeric states are indicated by a thick line. The previously known half-lives but remeasured in this work have been underlined by a red line, whereas the newly measured half-lives have been marked with a red box. The half-lives not measured in this work are also shown.}
\end{figure*}
\end{turnpage}

\section{\label{sec:Exp}Experimental Details}

The neutron-rich $^{122-131}$Sb isotopes were produced as fission fragments via fusion- and transfer-induced fission reactions at GANIL, using a $^{238}$U beam at an energy of 6.2 MeV/u, with typical beam intensity of 1 pnA on a $^{9}$Be target (1.6 and 5 $\mu$m thick). The unambiguous isotopic identification of the fission fragments ($Z$, $A$, $q$) was achieved using the VAMOS++ spectrometer, placed at 20$^{\circ}$ relative to the beam axis~\cite{re11, na14, nare}. The focal plane detection system of VAMOS++ was constituted of a Multi-Wire Proportional Counter (MWPC), two drift chambers and a segmented ionization chamber. The AGATA $\gamma$-ray tracking array, consisting of 32 crystals, was placed at 13.5 cm from the target position~\cite{ak12}. The velocity vector of the recoiling ions (measured by Dual Position-Sensitive MWPC (DPS-MWPC) detector~\cite{va16}, placed at the entrance of the VAMOS++ spectrometer, and the $\gamma$-ray emission angle (determined using AGATA) were used to obtain the Doppler corrected prompt $\gamma$ rays ($\gamma_{P}$), on an event-by-event basis. Gamma-ray interaction points, determined by Pulse Shape Analysis (PSA) and GRID search algorithm techniques, were tracked using the Orsay Forward Tracking (OFT) algorithm, as described in Ref.~\cite{ki17}. The delayed $\gamma$ rays ($\gamma_{D}$) were detected using seven EXOGAM HPGe Clover detectors~\cite{si00}, arranged in a wall like configuration at the focal plane of VAMOS++. The particle identification (PID) spectra, intrinsic resolutions of the spectrometer ($\Delta Z/Z$, $\Delta A/A$, $\Delta q/q$) and additional experimental details are given in Ref.~\cite{ki17}.

The contamination from neighboring isobars, arising from the resolution in atomic number $Z$, was subtracted for prompt $\gamma$-ray spectra. In addition to the background reduction for the delayed spectra as suggested in Ref.~\cite{ki17}, the subtraction in delayed spectra for neighboring isotopes was performed for all the Sb isotopes. The $\gamma$-$\gamma$ matrices ($\gamma_{P}-\gamma_{P}$, $\gamma_{D}-\gamma_{D}$, and $\gamma_{P}-~\gamma_{D}$) were generated for both the prompt (P) and delayed (D) transitions for all the Sb isotopes. The uncertainties in the energy of the prompt and delayed $\gamma$ rays is $\sim$ 1 keV. The efficiencies for the prompt and delayed $\gamma$ rays were determined separately, as mentioned in Ref.~\cite{ki17}. The correction for the half-lives as mentioned in Ref.~\cite{ki17}, was also taken into account. The spin-parities were  assigned based on systematics and shell model calculations. One- ($A_{0} \times e^{-t/A_{1}ln2} + A_{2}$), two- ($A_{0} \times \frac{A_{2}}{A_{1}-A_{2}} \times (e^{-t/A_{1}ln2}- e^{-t/A_{2}ln2})+ A_{3} \times e^{-t/A_{2}ln2} + A_{4}$) and three-component fits, where $A_{i}~(i = 0,...,4)$ are the fitting parameters, were carried out for the estimation of half-lives of the states, wherever required. The conversion coefficients used for the estimation of the B(E2) transition probabilities were taken from BrIcc~\cite{ki08}.

\section{\label{sec:Res}Experimental Results}

A summary of the level schemes for all the odd-A and even-A $^{122-131}$Sb isotopes are shown in Fig.~\ref{fig:sb_fig1}. Already known $\gamma$-ray transitions are shown in black. Newly identified prompt and delayed transitions are shown by red and blue, respectively. Already known half-lives which have not been measured in this work are shown. The remeasured and newly measured half-lives have been indicated by a red line and a red box, respectively. The width of the arrows are proportional to the intensity of the transitions. The intensities of the prompt and delayed transitions have been separately measured and the lowest transition is normalized to 100 in each case. The prompt $\gamma$-ray transition emitted by the complementary fission fragment~\cite{na14, nare} (mainly from the fusion-fission) are observed and identified (marked by "c" in all the spectra) in the low energy part of the $\gamma$-ray spectra. The random coincidences with X-ray emitted by $^{238}$U is also observed and is marked by an $@$ symbol. 

\subsection{\label{sec:122sb}$^{122}$Sb}

Previous measurements on $^{122}$Sb were reported in Refs.~\cite{gu77, la88}. The level scheme as obtained in the present work is shown in Fig.~\ref{fig:sb_fig1}. Table~\ref{tab:122sb_tab1} shows the properties of all the transitions assigned in this work.

\begin{table}[h]
\caption{\label{tab:122sb_tab1}Properties of the transitions assigned to $^{122}$Sb obtained in this work}
\resizebox{0.4\textwidth}{!}{
\begin{ruledtabular}
\begin{tabular}{ccccc}

 E$_{\gamma}$ & I$_{\gamma}$ & $J_{i}^{\pi} \rightarrow  J_{f}^{\pi}$ & $E_{i}$ & $E_{f}$\\ \hline\\
 445.2 & 100 & $(9^{-})\rightarrow$ (8)$^{-}$  & 445 & 0 \\
 657.2 & 63(14) & $(10^{-})\rightarrow$ (9$^{-})$  & 1102 & 445 \\
 1102.2 & 29(18) & $(10^{-})\rightarrow$ (8)$^{-}$  & 1102 & 0 \\ 

\end{tabular}
\end{ruledtabular}}
\end{table}

\begin{figure}[].
\includegraphics[width=1.0\columnwidth]{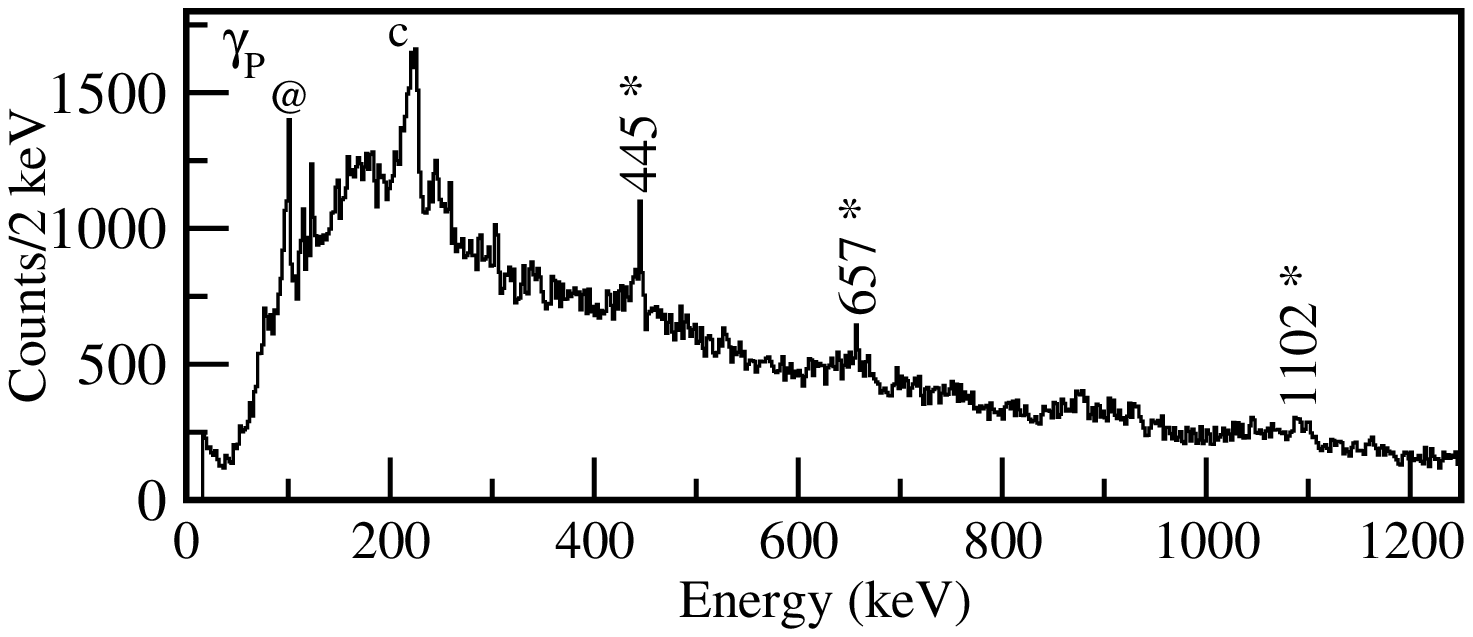}
\caption{\label{fig:122sb_fig2} $A-$ and $Z-$gated $\gamma$-ray spectrum for $^{122}$Sb: The tracked Doppler corrected prompt singles $\gamma$-ray ($\gamma_{P}$) spectrum. The newly identified $\gamma$ rays are marked with an asterisk.}
\end{figure}

The $A-$ and $Z-$gated tracked Doppler corrected prompt singles $\gamma$-ray ($\gamma_{P}$) spectrum for $^{122}$Sb is shown in Fig.~\ref{fig:122sb_fig2}. Three new prompt transitions 445, 657 and 1102~keV transitions are seen and these are marked with an asterisk. Based on the relative intensities, the 445~keV $\gamma$-ray is placed above the $(8)^{-}$ level (a 4.2 min isomeric state observed in Ref.~\cite{gu77}). In addition, it follows the systematics with the higher even-A Sb isotopes. Due to low statistics, no $\gamma_{P}$-$\gamma_{P}$ coincidence could be carried out. However, based on the summation of gammas, the intensities and the systematics, the 657~keV is placed above the 445~keV with the $(10^{-})$ level decaying by the 1102~keV to the $(8)^{-}$ state. No $A-$ and $Z-$gated delayed $\gamma$ rays ($\gamma_{D}$) in $^{122}$Sb could be identified from the present experiment.

\subsection{\label{sec:123sb}$^{123}$Sb}

\begin{figure}[t]
\includegraphics[width=1.0\columnwidth]{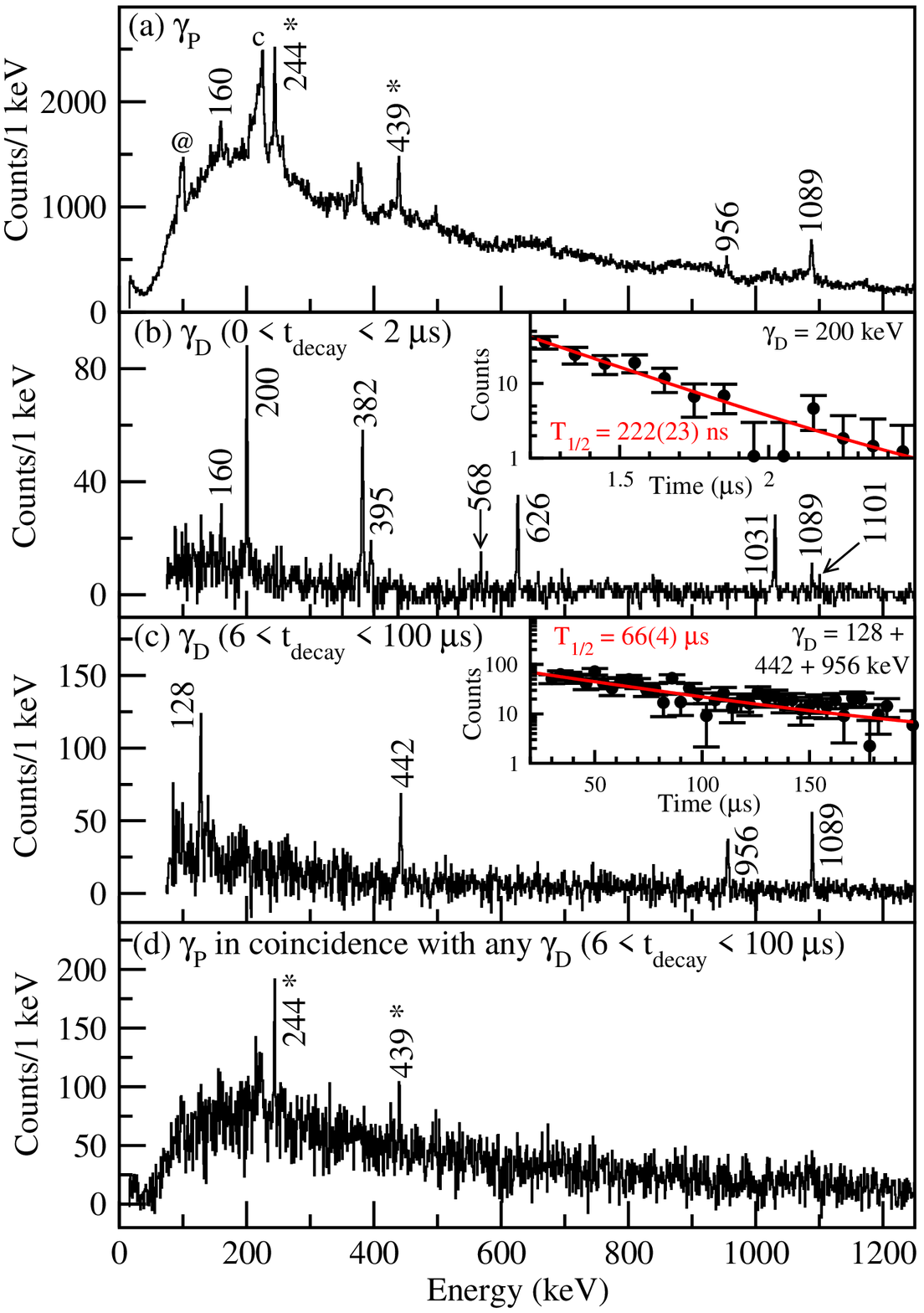}
\caption{\label{fig:123sb_fig2} (Color online) $A-$ and $Z-$gated $\gamma$-ray spectra for $^{123}$Sb: (a) The tracked Doppler corrected prompt $\gamma$-ray ($\gamma_{P}$) spectrum with the new $\gamma$-ray transitions marked with asterisk. (b) and (c) The delayed $\gamma$-ray ($\gamma_{D}$) spectra for 0 $< t_{decay} < $2~$\mu$s and 6 $< t_{decay} <$ 100~$\mu$s, respectively. The insets in (b) and (c) show the decay curves along with the fits for the 200 and 128 + 442 + 956~keV transitions, respectively. (d) $\gamma_{P}$ in coincidence with any $\gamma_{D}$ for 6 $< t_{decay} <$ 100~$\mu$s.}
\end{figure}

\begin{table}[b]
\caption{\label{tab:123sb_tab1}Properties of the transitions assigned to $^{123}$Sb obtained in this work. The top and bottom panels, separated by a line, are for the prompt and delayed transitions, respectively.}
\resizebox{0.4\textwidth}{!}{
\begin{ruledtabular}
\begin{tabular}{ccccc}

 E$_{\gamma}$ & I$_{\gamma}$ & $J_{i}^{\pi} \rightarrow  J_{f}^{\pi}$ & $E_{i}$ & $E_{f}$\\ \hline\\
 244.1 & 78(16) & (27/2$^{+})\rightarrow$ (25/2$^{+})$  & 3298 & 3054 \\
 439.2 & 100 & (25/2$^{+})\rightarrow$ 23/2$^{+}$  & 3054 & 2615 \\ \hline\\
 128.2 & 29(16) & 23/2$^{+}\rightarrow$ 19/2$^{+}$  & 2615 & 2487 \\
 159.8 & 63(36) & 5/2$^{+}\rightarrow$ 7/2$^{+}$  & 160 & 0 \\
 200.4 & 108(57) & (19/2$^{-})\rightarrow$ 15/2$^{-}$  & 2239 & 2039 \\
 382.0 & 112(54) & 15/2$^{-}\rightarrow$ 11/2$^{-}$  & 2039 & 1657 \\
 395.4 & 44(23) & 11/2$^{-}\rightarrow$ 9/2$^{+}$  & 1657 & 1261 \\
 441.6 & 32(11) & 19/2$^{+}\rightarrow$ 15/2$^{+}$  & 2487 & 2045 \\
 568.2 & 35(17) & 11/2$^{-}\rightarrow$ 11/2$^{+}$  & 1657 & 1089 \\
 626.0 & 125(56) & 11/2$^{-}\rightarrow$ 9/2$^{+}$  & 1657 & 1031 \\
 956.0 & 37(14) & 15/2$^{+}\rightarrow$ 11/2$^{+}$  & 2045 & 1089 \\
 1030.5 & 141(64) & 9/2$^{+}\rightarrow$ 7/2$^{+}$  & 1031 & 0 \\
 1089.0 & 100 & 11/2$^{+}\rightarrow$ 7/2$^{+}$  & 1089 & 0 \\
 1101.0 & 44(22) & 9/2$^{+}\rightarrow$ 5/2$^{+}$  & 1261 & 160 \\ 

\end{tabular}
\end{ruledtabular}}
\end{table}

The spectroscopy of  $^{123}$Sb isotope was previously studied in Refs.~\cite{po05, ju07, jo08, wa09}. The level scheme as obtained in the present work is shown in Fig.~\ref{fig:sb_fig1}. Table~\ref{tab:123sb_tab1} shows the properties of all the transitions assigned in this work.

The $A-$ and $Z-$gated $\gamma$-ray spectra for $^{123}$Sb are shown in Fig.~\ref{fig:123sb_fig2}. The tracked Doppler corrected prompt singles $\gamma$-ray spectrum ($\gamma_{P}$) for $^{123}$Sb is shown in Fig.~\ref{fig:123sb_fig2}(a). Previously observed 160, 956 and 1089~keV transitions are seen in this spectrum. Two new prompt $\gamma$-ray transitions, namely 244 and 439~keV, are identified and these are marked with an asterisk. The delayed $\gamma$-ray spectrum ($\gamma_{D}$) for 0~$< t_{decay} <$ 2~$\mu$s is shown in Fig.~\ref{fig:123sb_fig2}(b), yielding 160, 200, 382, 395, 568, 626, 1031, 1089 and 1101~keV transitions, as expected from previous works. The half-life fit (one-component) for the decay spectrum upon gating on 200~keV transition yields a value of $T_{1/2}$ = 222(23)~ns for the (19/2$^{-}$) state (in agreement with the quoted value of 214(3)~ns in Ref.~\cite{wa09}), which gives B(E2; $19/2^{-} \rightarrow 15/2^{-}$) = 6.9(7)~e$^{2}$fm$^{4}$. However, the weak delayed transitions, which were mentioned in Refs.~\cite{wa09, jo08}, namely the 949, 1007, 1260 and 1656~keV transitions, are not observed owing to the low statistics in our dataset. Similarly, Fig.~\ref{fig:123sb_fig2}(c) shows the delayed $\gamma$-ray spectrum ($\gamma_{D}$) for 6 $< t_{decay} <$ 100~$\mu$s, leading to 128, 442, 956 and 1089~keV transitions. A half-life fit (one-component) for the decay spectrum upon gating on 128, 442 and 956~keV transitions yields a value of $T_{1/2}$ = 66(4)~$\mu$s for the $23/2^{+}$ state (in agreement with the value of 65(1)~$\mu$s reported in Ref.~\cite{wa09}), which gives B(E2; $23/2^{+} \rightarrow 19/2^{+}$) = 0.15(1)~e$^{2}$fm$^{4}$. However, the weak transitions 100, 148, 348, 375 and 1250~keV, which were mentioned in Ref.~\cite{wa09}, are not observed owing to the low statistics in our dataset. The tracked Doppler corrected $\gamma_{P}$ in coincidence with any $\gamma_{D}$ (for 6 $< t_{decay} <$ 100~$\mu$s) is shown in Fig.~\ref{fig:123sb_fig2}(d). This spectrum yields the two newly identified prompt $\gamma$-ray transitions, 244 and 439~keV transitions. In order to confirm these newly found prompt transitions, a gate when applied on the prompt 244 and 439~keV transitions yielded the delayed 128~keV (not shown in this figure). Thus these two new transitions are placed above the $23/2^{+}$ isomer. However, the tracked Doppler corrected $\gamma_{P}$ in coincidence with any $\gamma_{D}$ for 0 $< t_{decay} <$ 2~$\mu$s, did not result in any new prompt $\gamma$ rays and hence no $\gamma$-ray is placed above the (19/2$^{-}$) isomer.

\subsection{\label{sec:124sb}$^{124}$Sb}

The $\gamma$-ray spectroscopy measurement on the high-spin states in $^{124}$Sb was previously reported in Ref.~\cite{re16}. The level scheme as obtained in the present work is shown in Fig.~\ref{fig:sb_fig1}. Table.~\ref{tab:124sb_tab1} shows the properties of all the transitions assigned in this work.

\begin{figure}[]
\includegraphics[width=1.0\columnwidth]{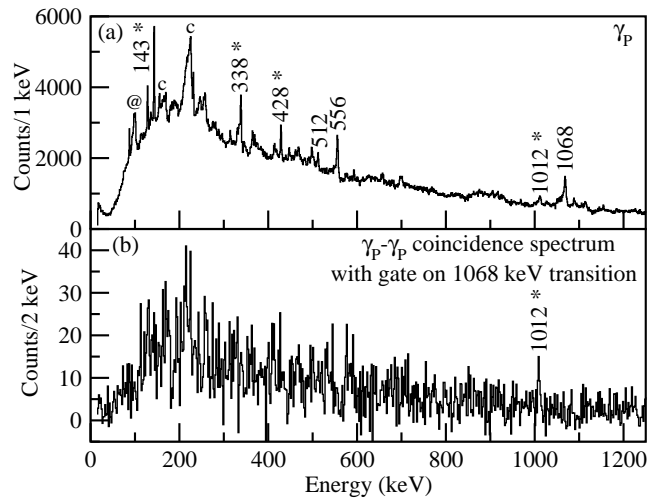}
\caption{\label{fig:124sb_fig2} $A-$ and $Z-$gated $\gamma$-ray spectra for $^{124}$Sb: (a) The tracked Doppler corrected prompt $\gamma$-ray ($\gamma_{P}$) singles spectrum. (b) The prompt $\gamma$-$\gamma$ ($\gamma_{P}$-$\gamma_{P}$) spectrum with gate on 1068~keV transition. The newly identified prompt transitions are marked with an asterisk.}
\end{figure}

The $A-$ and $Z-$gated tracked Doppler corrected prompt singles $\gamma$-ray spectrum for $^{124}$Sb is shown in Fig.~\ref{fig:124sb_fig2}(a). The already identified $\gamma$ rays in Ref.~\cite{re16}, namely 512, 556 and 1068~keV are seen. Other than these, four new prompt 143, 338, 428 and 1012~keV $\gamma$-ray transitions are identified (marked with an asterisk). Figure~\ref{fig:124sb_fig2}(b) shows the $A-$ and $Z-$gated tracked Doppler corrected prompt $\gamma$-$\gamma$ coincidence ($\gamma_{P}$-$\gamma_{P}$) spectrum with gate on the 1068~keV transition. This shows that the 1012~keV transition is in coincidence and hence is placed in the level scheme. The placement of 1012~keV transition in the level scheme follows the systematics with the other even-A Sb isotopes. The other prompt transitions 143, 338 and 428~keV are not placed as these are not observed in the coincidence spectrum. These $\gamma$ rays probably belong to the side band (as observed above the isomer in the case of $^{126}$Sb). No delayed $\gamma$ rays could be identified for $^{124}$Sb, using the present setup.

\begin{table}[h]
\caption{\label{tab:124sb_tab1}Properties of the transitions assigned to $^{124}$Sb obtained in this work}
\resizebox{0.4\textwidth}{!}{
\begin{ruledtabular}
\begin{tabular}{ccccc}

 E$_{\gamma}$ & I$_{\gamma}$ & $J_{i}^{\pi} \rightarrow  J_{f}^{\pi}$ & $E_{i}$ & $E_{f}$\\ \hline\\
 512.3 & 30(6) & ($10^{-})\rightarrow$ (9$^{-})$  & 1068 & 556 \\
 556.0 & 100 & ($9^{-})\rightarrow$ (8$^{-})$  & 556 & 0 \\
 1012.0 & 29(6) & (12$^{-})\rightarrow$ (10$^{-})$  & 2080 & 1068 \\
 1068.4 & 69(13) & (10$^{-})\rightarrow$ (8$^{-})$  & 1068 & 0 \\

\end{tabular}
\end{ruledtabular}}
\end{table}

\subsection{\label{sec:125sb}$^{125}$Sb}

\begin{figure}[]
\includegraphics[width=1.0\columnwidth]{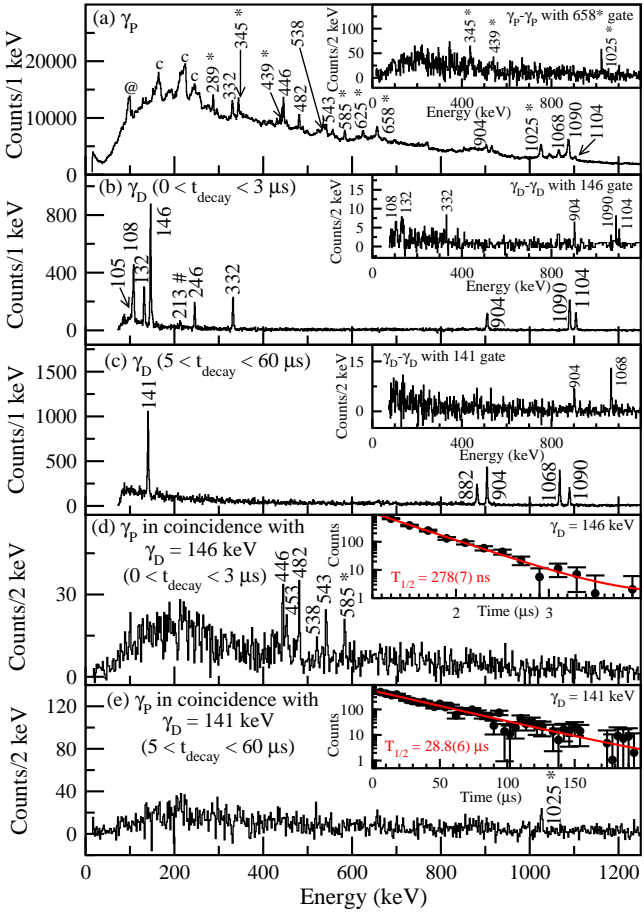}
\caption{\label{fig:125sb_fig2} (Color online) $A-$ and $Z-$gated $\gamma$-ray spectra for $^{125}$Sb: (a) The tracked Doppler corrected prompt singles $\gamma$-ray ($\gamma_{P}$) spectrum with the new $\gamma$-ray transitions marked with asterisk. The inset shows the tracked Doppler corrected prompt $\gamma_{P}$-$\gamma_{P}$ coincidence spectrum with gate on the newly observed 658~keV $\gamma$-ray (b) and (c) The delayed singles $\gamma$-ray ($\gamma_{D}$) spectra for 0 $<~t_{decay}~<$~3~$\mu$s and 5~$<~t_{decay}~<$~60~$\mu$s, respectively. The insets in (b) and (c) shows the delayed $\gamma_{D}$-$\gamma_{D}$ coincidence spectra with gate on 146 and 141~keV $\gamma$ rays, respectively. (d) Tracked Doppler corrected $\gamma_{P}$ in coincidence with $\gamma_{D}$ = 146~keV $\gamma$-ray (for 0~$<~t_{decay}~<$~3~$\mu$s). The inset shows the decay curve along with the fit for the 146~keV transition. (e) $\gamma_{P}$ in coincidence with the $\gamma_{D}$ = 141~keV $\gamma$-ray (for 5~$<~t_{decay}~<$~60~$\mu$s). The inset shows the decay curve along with the fit for the 141~keV transition.}
\end{figure}

Previous measurements on the $\gamma$-ray spectroscopy of high-spin states in $^{125}$Sb were reported in Refs.~\cite{po05, ju07, re10}. The level scheme as obtained in the present work is shown in Fig.~\ref{fig:sb_fig1}. Table~\ref{tab:125sb_tab1} shows the properties of all the transitions assigned in this work.

\begin{table}[]
\caption{\label{tab:125sb_tab1}Properties of the transitions assigned to $^{125}$Sb obtained in this work. The top and bottom panels, separated by a line, are for the prompt and delayed transitions, respectively.}
\resizebox{0.4\textwidth}{!}{
\begin{ruledtabular}
\begin{tabular}{ccccc}

 E$_{\gamma}$ & I$_{\gamma}$ & $J_{i}^{\pi} \rightarrow  J_{f}^{\pi}$ & $E_{i}$ & $E_{f}$\\ \hline\\
 345.3 & 45(5) & (31/2$^{-})\rightarrow$ (27/2$^{-})$  & 4138 & 3795 \\
 438.7 & 26(3) & (35/2$^{-})\rightarrow$ (31/2$^{-})$  & 4579 & 4138 \\
 446.5 & 100 & (25/2$^{+})\rightarrow$ 23/2$^{+}$  & 2919 & 2472 \\
 453.1 & 17(2) & (33/2$^{+})\rightarrow$ (31/2$^{+})$  & 4981 & 4528 \\
 482.0 & 50(3) & (27/2$^{+})\rightarrow$ (25/2$^{+})$  & 3401 & 2919 \\
 538.1 & 13(3) & (35/2$^{+})\rightarrow$ (33/2$^{+})$  & 5366 & 4934 \\
 542.7 & 30(2) & (29/2$^{+})\rightarrow$ (27/2$^{+})$  & 3943 & 3401 \\
 584.8 & 23(3) & (31/2$^{+})\rightarrow$ (29/2$^{+})$  & 4528 & 3943 \\
 657.8 & 79(6) & (27/2$^{-})\rightarrow$ (23/2$^{-})$  & 3795 & 3138 \\
 1025.4 & 100 & (23/2$^{-})\rightarrow$ 19/2$^{-}$  & 3138 & 2112 \\
\hline\\
 105.1 & 6(4) & 17/2$^{-}\rightarrow$ 19/2$^{-}$  & 2217 & 2112 \\
 108.1 & 29(19) & 19/2$^{+}\rightarrow$ 17/2$^{-}$  & 2326 & 2217 \\
 132.1 & 24(13) & 19/2$^{+}\rightarrow$ 15/2$^{+}$  & 2326 & 2194 \\
 140.7 & 51(28) & 19/2$^{-}\rightarrow$ 15/2$^{-}$  & 2112 & 1972 \\
 146.3 & 52(27) & 23/2$^{+}\rightarrow$ 19/2$^{+}$  & 2472 & 2326 \\
 213.3 & 5(2) & 19/2$^{+}\rightarrow$ 19/2$^{-}$  & 2326 & 2112 \\
 245.9 & 19(11) & 17/2$^{-}\rightarrow$ 15/2$^{-}$  & 2217 & 1972 \\
 332.0 & 34(13) & 19/2$^{+}\rightarrow$ 15/2$^{+}$  & 2326 & 1994 \\
 881.8 & 50(17) & 15/2$^{-}\rightarrow$ 11/2$^{+}$  & 1972 & 1090 \\
 903.8 & 95(32) & 15/2$^{-}\rightarrow$ 9/2$^{+}$  & 1972 & 1068 \\
 904.1 & 29(10) & 15/2$^{+}\rightarrow$ 11/2$^{+}$  & 1994 & 1090 \\
 1067.7 & 100 & 9/2$^{+}\rightarrow$ 7/2$^{+}$  & 1068 & 0 \\
 1089.7 & 54(18) & 11/2$^{+}\rightarrow$ 7/2$^{+}$  & 1090 & 0 \\
 1104.1 & 35(12) & 15/2$^{+}\rightarrow$ 11/2$^{+}$  & 2194 & 1090 \\ 

\end{tabular}
\end{ruledtabular}}
\end{table}

The $A-$ and $Z-$gated $\gamma$-ray spectra for $^{125}$Sb is shown in Fig.~\ref{fig:125sb_fig2}. The tracked Doppler corrected prompt singles $\gamma$-ray spectrum ($\gamma_{P}$) for $^{125}$Sb is shown in Fig.~\ref{fig:125sb_fig2}(a). The previously known 332, 446, 482, 453, 538, 543, 904, 1068, 1090 and 1104 keV transitions are observed. Seven new prompt $\gamma$-ray transitions, namely 289, 345, 439, 585, 625, 658, and 1025~keV transitions are identified (marked with an asterisk). The inset shows the tracked Doppler corrected prompt $\gamma_{P}$-$\gamma_{P}$ coincidence spectrum with gate on the newly identified 658 keV transition. This spectrum shows that the 345, 439, 658 and 1025~keV $\gamma$ rays are in coincidence. A similar coincidence spectrum was obtained for the 446, 453, 482, 538, 543 and 585~keV transitions (not shown in this figure). No $\gamma$ rays are seen in coincidence with the 289 and 625~keV transitions, hence these are not placed in the level scheme. The delayed $\gamma$-ray ($\gamma_{D}$) spectrum for 0~$<~t_{decay}~<$~3~$\mu$s is shown in Fig.~\ref{fig:125sb_fig2}(b) yielding 105, 108, 132, 146, 213, 246, 332, 904 (15/2$^{+} \rightarrow$ 11/2$^{+}$), 1090 and 1104~keV transitions, as expected from the level scheme. The 213~keV is newly identified and this is marked with an hash in the spectrum. The inset of Fig.~\ref{fig:125sb_fig2}(b) shows the delayed $\gamma_{D}$-$\gamma_{D}$ coincidence spectrum with gate on 146~keV transition, yielding all the intense delayed transitions. Similarly, Fig.~\ref{fig:125sb_fig2}(c) shows the delayed $\gamma$-ray spectrum for 5~$<~t_{decay}~<$~60~$\mu$s, leading to 141, 882, 904 (15/2$^{-} \rightarrow$ 9/2$^{+}$), 1068 and 1090~keV transitions. The inset of Fig.~\ref{fig:125sb_fig2}(c) shows the delayed $\gamma_{D}$-$\gamma_{D}$ coincidence spectrum with gate on 141~keV transition, yielding 904 and 1068~keV $\gamma$ rays as expected from previous works. The tracked Doppler corrected $\gamma_{P}$ in coincidence with $\gamma_{D}$ = 146~keV (for 0~$<~t_{decay}~<$~3~$\mu$s) is shown in Fig.~\ref{fig:125sb_fig2}(d). This spectrum yields the prompt $\gamma$-ray transitions, 446, 453, 482, 538, 543 and 585~keV transitions. In the previous Ref.~\cite{po05}, these prompt transitions, except 585~keV, were assigned negative-parities and placed above the $19/2^{-}$ isomer. Here we propose these transitions to be positive-parity states and placed above the $23/2^{+}$ isomer, as observed from prompt-delayed coincidences. Based on the intensities, the 585~keV transition is placed above the 543~keV transition, and the 453 and 538~keV transitions are replaced. The 432~keV transition, seen in Ref.~\cite{po05}, is not observed in coincidence, hence it is not assigned in the present level scheme. The half-life fit (one-component) for the decay spectrum upon gating on 146~keV transition yields a value of $T_{1/2}$~=~278(7)~ns for the $23/2^{+}$ (in agreement with the value of 272(16)~ns reported in Ref.~\cite{ju07}), which is shown in the inset of Fig.~\ref{fig:125sb_fig2}(d). This leads to B(E2; $23/2^{+} \rightarrow 19/2^{+}$) = 21.3(6)~e$^{2}$fm$^{4}$. The tracked Doppler corrected $\gamma_{P}$ in coincidence with $\gamma_{D}$ = 141~keV for 5~$<~t_{decay}~<$~60~$\mu$s, resulted in the observation of the newly identified 1025~keV $\gamma$-ray. As seen from the inset of Fig.~\ref{fig:125sb_fig2}(a), the 345, 439, 658 and 1025~keV transitions are in coincidence, and hence are placed above the $19/2^{-}$ isomer. The inset of Fig.~\ref{fig:125sb_fig2}(d) shows the half-life fit (two-component with one component fixed to 278 ns) for the decay spectrum upon gating on 141~keV transition yield a value of $T_{1/2}$ = 28.8(6)~$\mu$s for the $19/2^{-}$ state (in agreement with the quoted value of 28.0(7)~$\mu$s in Ref.~\cite{ju07}), which yields B(E2; $19/2^{-} \rightarrow 15/2^{-}$) = 0.24(1)~e$^{2}$fm$^{4}$.

\subsection{\label{sec:126sb}$^{126}$Sb}

\begin{figure}[h]
\includegraphics[width=1.0\columnwidth]{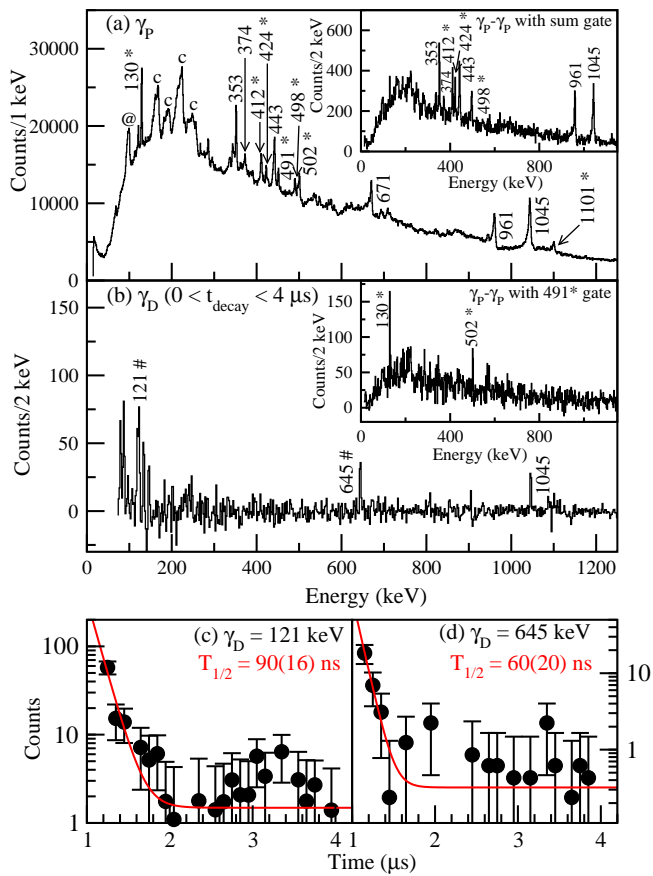}
\caption{\label{fig:126sb_fig2} (Color online) $A-$ and $Z-$gated $\gamma$-ray spectra for $^{126}$Sb: (a) Tracked Doppler corrected Prompt $\gamma$-ray singles spectrum ($\gamma_{P}$). The inset shows the tracked Doppler corrected $\gamma_{P}$-$\gamma_{P}$ coincidence spectrum with sum gate on $\gamma_{P}$'s, namely the 353, 443 and 961~keV transitions. The newly identified prompt $\gamma$ rays are marked with an asterisk. (b) The delayed singles $\gamma$-ray spectrum ($\gamma_{D}$) for 0~$<~t_{decay}~<$~4~$\mu$s. The newly identified delayed $\gamma$ rays are marked with hash. The inset shows the tracked Doppler corrected $\gamma_{P}$-$\gamma_{P}$ coincidence spectrum with gate on the newly identified 491~keV prompt $\gamma$-ray. (c) and (d) The decay curve along with the fit for the delayed 121 and 645~keV transitions, respectively.}
\end{figure}

\begin{table}[h]
\caption{\label{tab:126sb_tab1}Properties of the transitions assigned to $^{126}$Sb obtained in this work.  The top and bottom panels, separated by a line, are for the prompt and delayed transitions, respectively.}
\resizebox{0.4\textwidth}{!}{
\begin{ruledtabular}
\begin{tabular}{ccccc}

 E$_{\gamma}$ & I$_{\gamma}$ & $J_{i}^{\pi} \rightarrow  J_{f}^{\pi}$ & $E_{i}$ & $E_{f}$\\ \hline\\
 130.1 & 44(4) & (14$^{+})\rightarrow$ (13$^{+})$  & 2813 & 2682 \\
 352.9 & 72(7) & (14$^{-})\rightarrow$ (12$^{-})$  & 2359 & 2006 \\
 373.5 & 35(5) & (10$^{-})\rightarrow$ (9$^{-})$  & 1045 & 671 \\
 411.9 & 39(5) & (16$^{-})\rightarrow$ (15$^{-})$  & 3214 & 2802 \\
 423.5 & 17(2) & (17$^{-})\rightarrow$ (16$^{-})$  & 3638 & 3214 \\
 443.3 & 59(7) & (15$^{-})\rightarrow$ (14$^{-})$  & 2802 & 2359 \\
 491.4 & 78(7) & (13$^{+})\rightarrow$ (12$^{+})$  & 2682 & 2191 \\
 497.6 & 13(2) & (18$^{-})\rightarrow$ (17$^{-})$  & 4135 & 3638 \\
 501.8 & 100 & (12$^{+})\rightarrow$ (11$^{+})$  & 2191 & 1689 \\
 671.2 & 100 & (9$^{-})\rightarrow$ (8$^{-})$  & 671 & 0 \\ 
 961.4 & 79(11) & (12$^{-})\rightarrow$ (10$^{-})$  & 2006 & 1045 \\
 1044.9 & 87(7) & (10$^{-})\rightarrow$ (8$^{-})$  & 1045 & 0 \\ 
 1101.1 & 100 & (15$^{+})\rightarrow$ (13$^{+})$  & 2912 & 1811 \\ \hline\\
 121.3 & 70(43) & (13$^{+})\rightarrow$ (11$^{+})$  & 1811 & 1689 \\
 644.6 & 76(29) & (11$^{+})\rightarrow$ (10$^{-})$  & 1689 & 1045 \\ 

\end{tabular}
\end{ruledtabular}}
\end{table} 

Gamma-ray spectroscopy measurements on the high-spin states in $^{126}$Sb was previously reported in Ref.~\cite{re16}. The level scheme as obtained in the present work is shown in Fig.~\ref{fig:sb_fig1}. Table.~\ref{tab:126sb_tab1} shows the properties of all the transitions assigned in this work.

The $A-$ and $Z-$gated $\gamma$-ray spectra for $^{126}$Sb are shown in Fig.~\ref{fig:126sb_fig2}. The tracked Doppler corrected prompt singles $\gamma$-ray spectrum ($\gamma_{P}$) for $^{126}$Sb is shown in Fig.~\ref{fig:126sb_fig2}(a). The already known $\gamma$ rays from Ref.~\cite{re16}, namely 353, 374, 443, 671, 961 and 1045~keV are observed. In addition, new prompt $\gamma$ rays 130, 412, 424, 491, 498, 502 and 1101~keV (marked with an asterisk) are identified. The tracked Doppler corrected prompt $\gamma$-$\gamma$ coincidence ($\gamma_{P}$-$\gamma_{P}$) spectrum with a sum gate of 353, 443 and 961~keV prompt $\gamma$-ray transitions is shown in the inset of Fig.~\ref{fig:126sb_fig2}(a). This spectrum shows that the 353, 374, 412, 424, 443, 498, 671, 961 and 1045~keV transitions are in coincidence and the newly observed transitions are marked with an asterisk. Figure~\ref{fig:126sb_fig2}(b) shows the delayed $\gamma$-ray ($\gamma_{D}$) singles spectrum, with 0~$<~t_{decay}~<$~4~$\mu$s. This spectrum shows the presence of two new delayed 121 and 645~keV transitions, besides the 1045~keV transition (marked with an hash). The inset in Fig.~\ref{fig:126sb_fig2}(b) shows the tracked Doppler corrected prompt $\gamma$-$\gamma$ coincidence spectrum ($\gamma_{P}$-$\gamma_{P}$) gated on the newly identified 491~keV transition. This shows that the 130, 491 and 502~keV transitions are in coincidence and these are placed above the $(11^{+})$ state in the level scheme, following the systematics with higher even-A Sb isotopes. Also in Fig.~\ref{fig:126sb_fig2}(a), the newly identified 1101~keV transition, is not observed in the tracked Doppler corrected prompt $\gamma_{P}-\gamma_{P}$ spectra (insets of Fig.~\ref{fig:126sb_fig2}(a) and (b)). Following the systematics with $^{128}$Sb, this is placed above the $(13^{+})$ state. The decay spectrum of the delayed 121~keV transition is shown in Fig.~\ref{fig:126sb_fig2}(c). An exponential fit (one-component) yields a value of 90(16)~ns. This gives B(E2; $13^{+} \rightarrow 11^{+}$) = 131(28)~e$^{2}$fm$^{4}$. The half-life of the 645~keV transition was also measured with $T_{1/2}$ = 60(20)~ns, as shown in Fig.~\ref{fig:126sb_fig2}(d) (using a two-component fit with one component fixed to 90 ns).

\subsection{\label{sec:127sb}$^{127}$Sb}

Previous $\gamma$-ray spectroscopy measurements on $^{127}$Sb were reported in Refs.~\cite{ap74, po05, wa09epja}. The level scheme as obtained in the present work is shown in Fig.~\ref{fig:sb_fig1}. Table~\ref{tab:127sb_tab1} shows the properties of all the transitions assigned in this work. 

\begin{figure}[t]
\includegraphics[width=1.0\columnwidth]{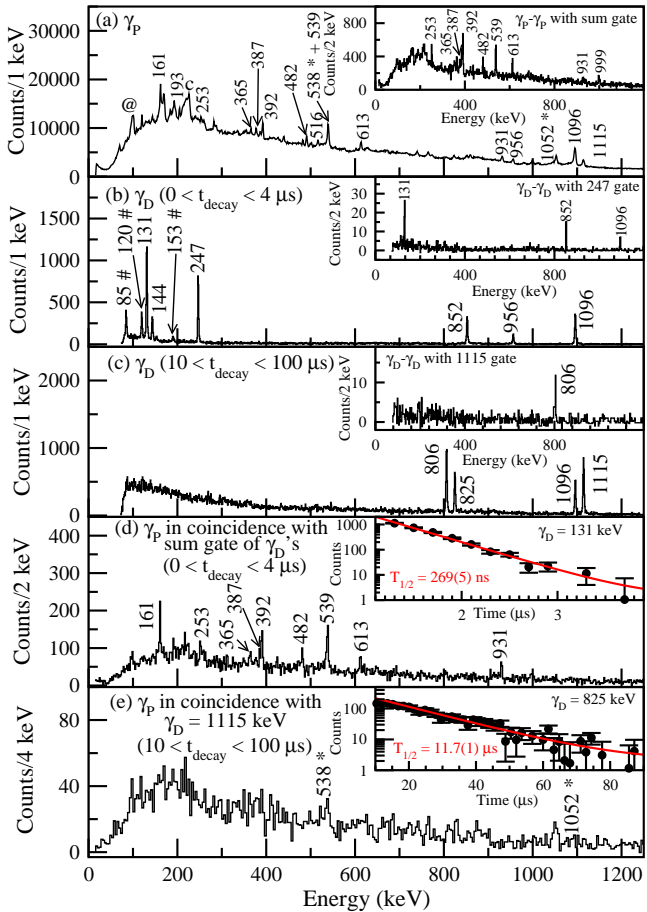}
\caption{\label{fig:127sb_fig2} (Color online) $A-$ and $Z-$gated $\gamma$-ray spectra for $^{127}$Sb: (a) The tracked Doppler corrected prompt singles $\gamma$-ray ($\gamma_{P}$) spectrum with the new $\gamma$-ray transitions marked with asterisk. The inset shows the prompt $\gamma_{P}$-$\gamma_{P}$ coincidence spectrum with a sum gate on $\gamma_{P}$'s, namely 387, 392, 539 and 613~keV $\gamma$ rays. (b) and (c) The delayed singles $\gamma$-ray ($\gamma_{D}$) spectra for 0~$<~t_{decay}~<$~4~$\mu$s and 10~$<~t_{decay}~<$~100~$\mu$s, respectively. The insets in (b) and (c) shows the delayed $\gamma_{D}$-$\gamma_{D}$ coincidence spectra with gate on 247 and 1115~keV $\gamma$ rays. (d) Tracked Doppler corrected $\gamma_{P}$ in coincidence with the sum gate of the $\gamma_{D}$'s, namely 85, 120, 131, 144, 247, 852, 956 and 1096~keV $\gamma$ rays (for 0~$<~t_{decay}~<$~4~$\mu$s). The inset shows the decay curve along ith the fit for the 131~keV transition. (e) Tracked Doppler corrected $\gamma_{P}$ in coincidence with the $\gamma_{D}$ = 1115~keV $\gamma$-ray (for 10~$<~t_{decay}~<$~100~$\mu$s). The inset shows the decay curve along with the fit for the 825~keV transition.}
\end{figure} 

\begin{table}[h]
\caption{\label{tab:127sb_tab1}Properties of the transitions assigned to $^{127}$Sb obtained in this work.  The top and bottom panels, separated by a line, are for the prompt and delayed transitions, respectively.}
\resizebox{0.4\textwidth}{!}{
\begin{ruledtabular}
\begin{tabular}{ccccc}

 E$_{\gamma}$ & I$_{\gamma}$ & $J_{i}^{\pi} \rightarrow  J_{f}^{\pi}$ & $E_{i}$ & $E_{f}$\\ \hline\\
  161.4 & 38(2) & (21/2$^{+})\rightarrow$ 23/2$^{+}$  & 2486 & 2325 \\
 192.6 & 15(2) & (23/2$^{+})\rightarrow$ (21/2$^{+})$  & 2678 & 2486 \\
 252.7 & 11(5) & (37/2$^{+})\rightarrow$ (35/2$^{+})$  & 5354 & 5101 \\
 365.1 & 12(2) & (35/2$^{+})\rightarrow$ (33/2$^{+})$  & 5101 & 4736 \\
 386.7 & 31(3) & (31/2$^{+})\rightarrow$ (29/2$^{+})$  & 4255 & 3868 \\
 391.8 & 60(3) & (27/2$^{+})\rightarrow$ (25/2$^{+})$  & 3256 & 2864 \\
 482.0 & 17(5) & (33/2$^{+})\rightarrow$ (31/2$^{+})$  & 4736 & 4255 \\
 516.1 & 13(3) & (25/2$^{+})\rightarrow$ (23/2$^{+})$  & 3194 & 2678 \\
 538.1 & 30(2) & (27/2$^{-})\rightarrow$ (23/2$^{-})$  & 3579 & 3041 \\
 539.1 & 100 & (25/2$^{+})\rightarrow$ 23/2$^{+}$  & 2864 & 2325 \\
 612.8 & 46(3) & (29/2$^{+})\rightarrow$ (27/2$^{+})$  & 3868 & 3256 \\
 930.9 & 32(2) & (27/2$^{+})\rightarrow$ (23/2$^{+})$  & 3256 & 2325 \\
 999.0 & 21(3) & (31/2$^{+})\rightarrow$ (27/2$^{+})$  & 4255 & 3256 \\
 1052.4 & 100 & (23/2$^{-})\rightarrow$ (19/2$^{-})$  & 3041 & 1989 \\ \hline\\
 84.8 & 11(9) & (17/2$^{-})\rightarrow$ (19/2$^{-})$  & 2074 & 1989 \\
 119.9 & 12(7) & 19/2$^{+}\rightarrow$ (17/2$^{-})$  & 2194 & 2074 \\
 131.1 & 49(28) & 23/2$^{+}\rightarrow$ 19/2$^{+}$  & 2325 & 2194 \\
 143.5 & 14(7) & 19/2$^{+}\rightarrow$ 15/2$^{+}$  & 2194 & 2051 \\
 153.1 & 2(1) & (17/2$^{-})\rightarrow$ 15/2$^{-}$  & 2074 & 1920 \\
 246.9 & 47(20) & 19/2$^{+}\rightarrow$ 15/2$^{+}$  & 2194 & 1948 \\
 806.0 & 97(33) & 15/2$^{-}\rightarrow$ 9/2$^{+}$  & 1920 & 1115 \\
 824.8 & 61(21) & 15/2$^{-}\rightarrow$ 11/2$^{+}$  & 1920 & 1096 \\
 852.5 & 43(15) & 15/2$^{+}\rightarrow$ 11/2$^{+}$  & 1948 & 1096 \\
 956.2 & 17(6) & 15/2$^{+}\rightarrow$ 11/2$^{+}$  & 2051 & 1096 \\
 1095.8 & 61(20) & 11/2$^{-}\rightarrow$ 7/2$^{+}$  & 1096 & 0 \\
 1114.6 & 100 & 9/2$^{+}\rightarrow$ 7/2$^{+}$  & 1115 & 0 \\

\end{tabular}
\end{ruledtabular}}
\end{table}

The $A-$ and $Z-$gated $\gamma$-ray spectra for $^{127}$Sb is shown in Fig.~\ref{fig:127sb_fig2}. The tracked Doppler corrected prompt singles $\gamma$-ray spectrum ($\gamma_{P}$) for $^{127}$Sb is shown in Fig.~\ref{fig:127sb_fig2}(a). The previously known 161, 193, 253, 365, 387, 392, 482, 516, 613, 931, 956, 999, 1096 and 1115 keV transitions are seen. Two new prompt $\gamma$-ray transitions, namely 538 and 1052~keV transitions, are identified and these are marked with an asterisk. The inset shows the tracked Doppler corrected prompt $\gamma_{P}$-$\gamma_{P}$ coincidence spectrum with sum gate on the 387, 392, 539, and 613~keV $\gamma$-ray transitions. This spectrum shows that the 253, 365, 387, 392, 482, 539, 613, 931 and 999~keV $\gamma$ rays are in coincidence. A similar coincidence spectrum is obtained for 161, 193, and 516~keV transitions (not shown in this figure). The delayed singles $\gamma$-ray ($\gamma_{D}$) spectrum for 0~$<~t_{decay}~<$~4~$\mu$s is shown in Fig.~\ref{fig:127sb_fig2}(b), yielding 85, 120, 131, 144, 153, 247, 852, 956 and 1096~keV transitions, as expected from the previous works. The 85, 120, and 153~keV transitions are newly identified (marked with hash). The inset shows the delayed $\gamma_{D}$-$\gamma_{D}$ coincidence spectrum with gate on 247~keV transition, yielding 131, 852 and 1096~keV transitions, as reported in previous measurements. The newly observed delayed transitions are placed in accordance with that of $^{125}$Sb and summation of gammas. Similarly, Fig.~\ref{fig:127sb_fig2}(c) shows the singles $\gamma_{D}$-ray spectrum for 10~$<~t_{decay}~<$~100~$\mu$s, leading to 806, 825, 1096 and 1115~keV transitions. The inset shows the delayed $\gamma_{D}$-$\gamma_{D}$ coincidence spectrum with gate on 1115~keV transition, yielding 806~keV $\gamma$-ray as expected from previous works. The $\gamma$-ray transition, 69~keV, between the ($19/2^{-}$) and $15/2^{-}$ (similar 200 and 141~keV transitions were observed in $^{123,125}$Sb isotopes, respectively) could not be observed with the present setup, as it is below the energy threshold. Assuming similar structure with the other odd-A Sb isotopes, the ($19/2^{-}$) state can be assumed to be an isomeric state in $^{127}$Sb. The tracked Doppler corrected $\gamma_{P}$ in coincidence with sum gate of $\gamma_{D}$'s namely 85, 120, 131, 144, 247, 852, 956, and 1096~keV transitions (for 0~$<~t_{decay}~<$~4~$\mu$s) is shown in Fig.~\ref{fig:127sb_fig2}(d). This spectrum yields the prompt $\gamma$-ray transitions, 161, 253, 365, 387, 392, 482, 539, 613 and 931~keV transitions. In the previous reference~\cite{po05}, these prompt transitions were assigned negative-parities and placed above the ($19/2^{-}$) isomer. Instead, we suggest these to be positive-parity states (placed above the $23/2^{+}$ isomer), as observed from the prompt-delayed coincidences. The half-life fit (one-component), shown in the inset of Fig.~\ref{fig:127sb_fig2}(c), for the decay spectrum upon gating on 131~keV transition yields a value of $T_{1/2}$ = 269(5)~ns for the $23/2^{+}$ state (close to the value of 234(12)~ns reported in Ref.~\cite{wa09epja}), which has been shown in the inset. This gives B(E2; $23/2^{+} \rightarrow 19/2^{+}$) = 33.4(7)~e$^{2}$fm$^{4}$. However, the tracked Doppler corrected $\gamma_{P}$ in coincidence with $\gamma_{D}$ = 1115~keV for 10~$<~t_{decay}~<$~100~$\mu$s, resulted in the observation of the newly identified 538 and 1052~keV $\gamma$ rays. These two transitions are thus placed above the ($19/2^{-}$) isomer, in accordance with the other odd-A Sb isotopes. The inset shows the half-life fit (two-component with one component fixed to 269 ns) for the decay spectrum upon gating on 825~keV transition, which yields a value of $T_{1/2}$ = 11.7(1)~$\mu$s for the $15/2^{-}$ isomer (in agreement with the value of 11(1)~$\mu$s reported in Ref.~\cite{ap74}).

\subsection{\label{sec:128sb}$^{128}$Sb}

\begin{figure}[t]
\includegraphics[width=1.0\columnwidth]{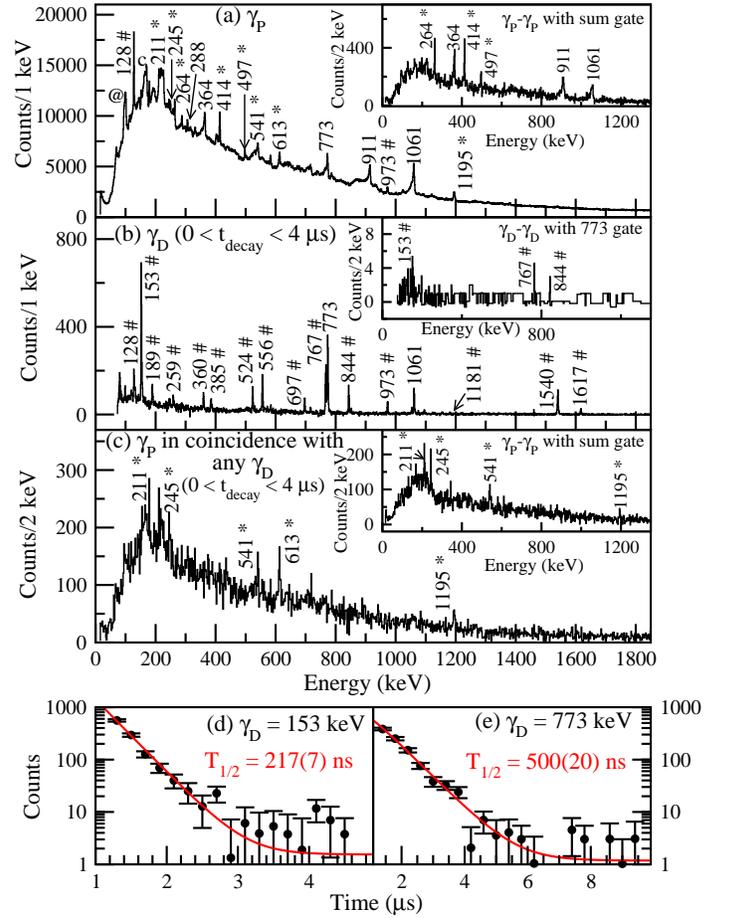}
\caption{\label{fig:128sb_fig2} (Color online) $A-$ and $Z-$gated $\gamma$-ray spectra for $^{128}$Sb: (a) Tracked Doppler corrected Prompt singles $\gamma$-ray spectrum ($\gamma_{P}$). The newly identified prompt and delayed $\gamma$ rays are marked with an asterisk and hash, respectively. The inset shows the $\gamma_{P}$-$\gamma_{P}$ coincidence spectrum with sum gate on the $\gamma_{P}$'s, namely the 264, 364 and 414~keV transitions. (b) The delayed singles $\gamma$-ray spectrum ($\gamma_{D}$) for 0~$<~t_{decay}~<$~4~$\mu$s. The newly identified delayed $\gamma$ rays are marked with hash. The inset shows the $\gamma_{D}$-$\gamma_{D}$ coincidence spectrum with gate on the delayed 773~keV $\gamma$-ray. (c) The $\gamma_{P}$ in coincidence with any delayed $\gamma$-ray for 0~$<~t_{decay}~<$~4~$\mu$s. Again, the newly identified prompt transitions are marked with an asterisk. The inset shows the $\gamma_{P}$-$\gamma_{P}$ coincidence spectrum with sum gate on many $\gamma_{P}$'s, namely the 541 and 1195~keV transitions. (d) and (e) shows the decay curves along with the fits for the delayed 153 and 773~keV transitions, respectively.}
\end{figure}

\begin{table}[b]
\caption{\label{tab:128sb_tab1}Properties of the transitions assigned to $^{128}$Sb obtained in this work.  The top and bottom panels, separated by a line, are for the prompt and delayed transitions, respectively.}
\resizebox{0.4\textwidth}{!}{
\begin{ruledtabular}
\begin{tabular}{ccccc}

 E$_{\gamma}$ & I$_{\gamma}$ & $J_{i}^{\pi} \rightarrow  J_{f}^{\pi}$ & $E_{i}$ & $E_{f}$\\ \hline\\
 211.1 & 10(5) & (18$^{+})\rightarrow$ (17$^{+})$  & 3961 & 3749 \\
 244.8 & 20(4) & (17$^{+})\rightarrow$ (16$^{+})$  & 3749 & 3505 \\
 263.9 & 37(4) & (16$^{-})\rightarrow$ (15$^{-})$  & 3011 & 2748 \\
 288.4 & 27(3) & (10$^{-})\rightarrow$ (9$^{-})$  & 1061 & 773 \\
 364.3 & 60(1) & (14$^{-})\rightarrow$ (12$^{-})$  & 2333 & 1969 \\
 413.9 & 52(2) & (15$^{-})\rightarrow$ (14$^{-})$  & 2748 & 2333 \\
 497.0 & 20(1) & (17$^{-})\rightarrow$ (16$^{-})$  & 3508 & 3011 \\
 541.3 & 58(3) & (16$^{+})\rightarrow$ (15$^{+})$  & 3505 & 2964 \\
 612.9 & 100 & (12$^{+})\rightarrow$ (11$^{+})$  & 2230 & 1617 \\
 773.0 & 100 & (9$^{-})\rightarrow$ (8$^{-})$  & 773 & 0 \\ 
 910.9 & 76(2) & (12$^{-})\rightarrow$ (10$^{-})$  & 1969 & 1061 \\
 1061.4 & 46(2) & (10$^{-})\rightarrow$ (8$^{-})$  & 1061 & 0 \\ 
 1195.1 & 100 & (15$^{+})\rightarrow$ (13$^{+})$  & 2964 & 1769 \\ \hline\\
 127.8 & 12(7) & (9$^{+})\rightarrow$ (8$^{+})$  & 1540 & 1412 \\
 152.6 & 45(24) & (13$^{+})\rightarrow$ (11$^{+})$  & 1769 & 1617 \\
 189.4 & 11(5) & (8$^{+})\rightarrow$ (7$^{+})$  & 1412 & 1223 \\
 259.4 & 11(5) & (7$^{-})\rightarrow$ (8$^{-})$  & 259 & 0 \\
 359.9 & 14(6) & (7$^{+})\rightarrow$ (8$^{-})$  & 360 & 0 \\
 384.8 & 13(5) & (11$^{+})\rightarrow$ (9$^{-})$  & 1617 & 1232 \\
 523.6 & 25(9) & (7$^{+})\rightarrow$ (6$^{+})$  & 1223 & 697 \\
 556.2 & 42(15) & (11$^{+})\rightarrow$ (10$^{-})$  & 1617 & 1061 \\
 696.9 & 17(6) & (6$^{+})\rightarrow$ (8$^{-})$  & 697 & 0 \\
 767.0 & 63(22) & (9$^{+})\rightarrow$ (9$^{-})$  & 1540 & 773 \\ 
 843.8 & 32(11) & (11$^{+})\rightarrow$ (9$^{-})$  & 1617 & 773 \\
 973.3 & 20(7) & (9$^{-})\rightarrow$ (7$^{-})$  & 1232 & 259 \\ 
 1181.2 & 4(2) & (9$^{+})\rightarrow$ (7$^{-})$  & 1540 & 360 \\
 1540.4 & 43(15) & (9$^{+})\rightarrow$ (8$^{-})$  & 1540 & 0 \\ 
 1617.2 & 10(3) & (11$^{+})\rightarrow$ (8$^{-})$  & 1617 & 0 \\ 

\end{tabular}
\end{ruledtabular}}
\end{table}

Previous $\gamma$-ray spectroscopy measurement on the high-spin states in $^{128}$Sb was shown in Ref.~\cite{re16}. The level scheme as obtained in the present work is shown in Fig.~\ref{fig:sb_fig1}. Table.~\ref{tab:128sb_tab1} shows the properties of all the transitions assigned in this work.

The $A-$ and $Z-$gated $\gamma$-ray spectra for $^{128}$Sb is shown in Fig.~\ref{fig:128sb_fig2}. The tracked Doppler corrected prompt singles $\gamma$-ray spectrum ($\gamma_{P}$) for $^{128}$Sb is shown in Fig.~\ref{fig:128sb_fig2}(a). The already known $\gamma$-rays, namely 288, 364, 773, 911, and 1061 keV, are seen. In addition, this spectrum shows many new prompt $\gamma$ rays, namely the 211, 245, 264, 414, 497, 541, 613 and 1195~keV transitions (marked with asterisk). Besides, the 128 and 973~keV transitions are also seen (marked with hash) are observed in both $\gamma_{P}$ (Fig.~\ref{fig:128sb_fig2}(a)) and delayed spectrum ($\gamma_{D}$) (Fig.~\ref{fig:128sb_fig2}(b)). The inset in Fig.~\ref{fig:128sb_fig2}(a) shows the prompt $\gamma$-$\gamma$ coincidence spectrum ($\gamma_{P}$-$\gamma_{P}$) with a sum gate on the 264, 364 and 414~keV prompt transitions. This spectrum shows that the 264, 364, 414, 497, 911 and 1061~keV transitions are in coincidence. The delayed $\gamma$-ray ($\gamma_{D}$) singles spectrum, with 0~$<~t_{decay}~<$~4~$\mu$s is shown in Fig.~\ref{fig:128sb_fig2}(b). This spectrum shows new delayed $\gamma$-ray transitions namely, 128, 153, 189, 259, 360, 385, 524, 556, 697, 767, 844, 973, 1181, 1540 and 1617~keV (marked with hash), in addition to the known 773 and 1061 keV transition. The delayed $\gamma$-$\gamma$ coincidence spectrum ($\gamma_{D}$-$\gamma_{D}$) with gate on the 773~keV transition is shown in the inset of Fig.~\ref{fig:128sb_fig2}(b). This spectrum shows that the 153, 767 and 844~keV are in coincidence with the 773~keV transition. Figure~\ref{fig:128sb_fig2}(c) shows the prompt spectrum ($\gamma_{P}$) in coincidence with any delayed $\gamma$-ray transition with 0~$<~t_{decay}~<$~4~$\mu$s is shown in Fig.~\ref{fig:128sb_fig2}(c). The inset in Fig.~\ref{fig:128sb_fig2}(c) shows the $\gamma_{P}$-$\gamma_{P}$ coincidence spectrum with gate on the newly identified 541 and 1195~keV transitions. This spectrum shows that 211, 245, 541 and 1195~keV transitions are in coincidence and placed above the ($13^{+}$) isomer. But Fig.~\ref{fig:128sb_fig2}(c) shows 613~keV transition which is not present in the inset of Fig.~\ref{fig:128sb_fig2}(c). Hence, the 613~keV transition is placed above the ($11^{+}$) isomer. Fig.~\ref{fig:128sb_fig2}(d) and (e) shows the decay curves for 153 and 773~keV transitions. An exponential fit for these curves yields values 217(7)~ns (using one-component fit) and 500(20)~ns (using two-component fit with one component fixed to 217 ns), respectively. This gives B(E2;~$13^{+} \rightarrow 11^{+}$)~=~23.1(8)~e$^{2}$fm$^{4}$.

\subsection{\label{sec:129sb}$^{129}$Sb}

References ~\cite{ge03, hu82, st87} reported the previous $\gamma$-ray spectroscopy measurements on $^{129}$Sb. The level scheme as obtained in the present work is shown in Fig.~\ref{fig:sb_fig1}. Table~\ref{tab:129sb_tab1} shows the properties of all the transitions assigned in this work.

\begin{figure}[h]
\includegraphics[width=1.0\columnwidth]{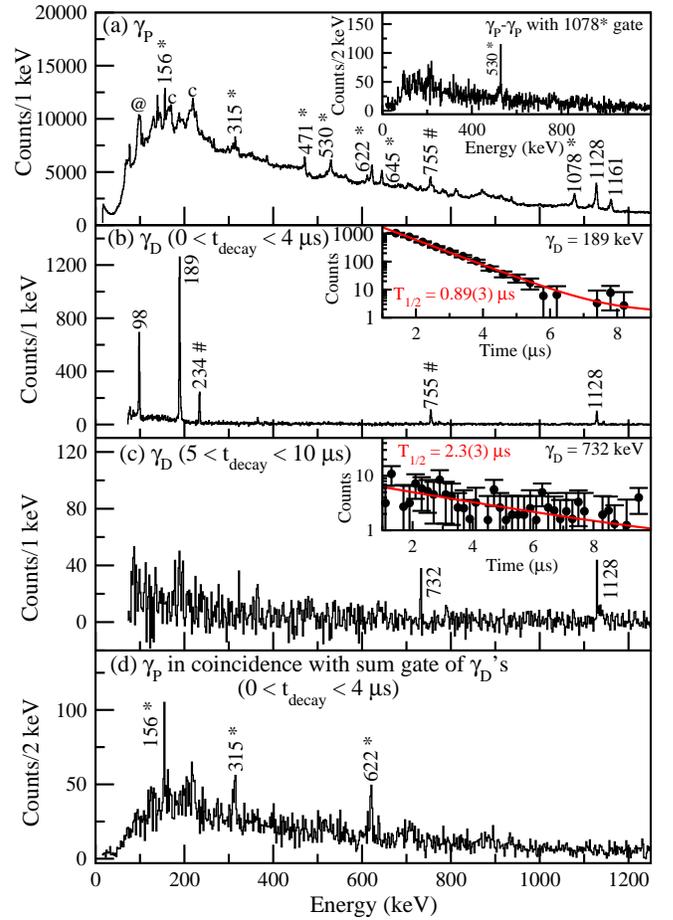}
\caption{\label{fig:129sb_fig2} (Color online) $A-$ and $Z-$gated $\gamma$-ray spectra for $^{129}$Sb: (a) The tracked Doppler corrected prompt singles $\gamma$-ray ($\gamma_{P}$) spectrum with the new $\gamma$-ray transitions marked with asterisk. The inset shows the tracked Doppler corrected prompt $\gamma_{P}$-$\gamma_{P}$ coincidence spectrum with gate on the newly identified 1078~keV transition. (b) and (c) The delayed singles $\gamma$-ray ($\gamma_{D}$) spectra for 0~$<~t_{decay}~<$~4~$\mu$s and 5~$<~t_{decay}~<$~10 $\mu$s, respectively. The insets in (b) and (c) shows the decay curves along with the fits for the 189 and 732~keV transitions, respectively. (d) Tracked Doppler corrected $\gamma_{P}$ in coincidence with the sum gate of the $\gamma_{D}$'s, namely 98, 189, 234 and 755~keV delayed transitions (for 0~$<~t_{decay}~<$~4~$\mu$s).}
\end{figure}

\begin{table}[b]
\caption{\label{tab:129sb_tab1}Properties of the transitions assigned to $^{129}$Sb obtained in this work.  The top and bottom panels, separated by a line, are for the prompt and delayed transitions, respectively.}
\resizebox{0.4\textwidth}{!}{
\begin{ruledtabular}
\begin{tabular}{ccccc}

 E$_{\gamma}$ & I$_{\gamma}$ & $J_{i}^{\pi} \rightarrow  J_{f}^{\pi}$ & $E_{i}$ & $E_{f}$\\ \hline\\
 156.1 & 55(4) & (21/2$^{+})\rightarrow$ 23/2$^{+}$  & 2295 & 2139 \\
 314.7 & 44(4) & (27/2$^{+})\rightarrow$ (25/2$^{+})$  & 3075 & 2761 \\
 529.7 & 69(10) & (27/2$^{-})\rightarrow$ (23/2$^{-})$  & 3469 & 2939 \\
 622.1 & 100 & (25/2$^{+})\rightarrow$ 23/2$^{+}$  & 2761 & 2139 \\
 1078.6 & 100 & (23/2$^{-})\rightarrow$ 19/2$^{-}$  & 2939 & 1851 \\ \hline\\
 98.3 & 38(7) & 23/2$^{+}\rightarrow$ 19/2$^{+}$  & 2139 & 2040 \\
 189.2 & 71(34) & 19/2$^{+}\rightarrow$ 19/2$^{-}$  & 2040 & 1851 \\
 234.0 & 35(16) & 19/2$^{+}\rightarrow$ (17/2$^{-})$  & 2040 & 1806 \\
 732.5 & 47(17) & 15/2$^{-}\rightarrow$ 11/2$^{+}$  & 1861 & 1128 \\
 755.0 & 37(15) & (15/2$^{+})\rightarrow$ 11/2$^{+}$  & 1883 & 1128 \\
 1128.5 & 100 & 11/2$^{+}\rightarrow$ 7/2$^{+}$  & 1128 & 0 \\ 

\end{tabular}
\end{ruledtabular}}
\end{table}

The $A-$ and $Z-$gated $\gamma$-ray spectra for $^{129}$Sb are shown in Fig.~\ref{fig:129sb_fig2}. The tracked Doppler corrected prompt singles $\gamma$-ray spectrum ($\gamma_{P}$) for $^{129}$Sb is shown in Fig.~\ref{fig:129sb_fig2}(a). The previously observed 1128 and 1161 keV transitions are seen. Many new prompt $\gamma$-ray transitions, namely 156, 315, 471, 530, 622, 645 and 1078~keV transitions are identified (marked with an asterisk). The 755~keV (marked with an hash) is observed in  both prompt ($\gamma_{P}$) (Fig.~\ref{fig:129sb_fig2}(a)) and delayed spectrum ($\gamma_{D}$) (Fig.~\ref{fig:129sb_fig2}(b)). The 1161~keV is actually depopulating the $9/2^{+}$ state to the ground-state~\cite{hu82, st87}, and this is not shown in the current level scheme. The inset shows the tracked Doppler corrected prompt $\gamma_{P}$-$\gamma_{P}$ coincidence spectrum with gate on the newly identified 1078~keV transition. This spectrum shows that the 530 and 1078~keV transitions are in coincidence. A similar coincidence spectrum is obtained for the 315 and 622~keV transitions (not shown in this figure). However, no coincidences are seen upon gating on the 471 and 645~keV transitions, and hence these are not placed in the level scheme. The delayed singles $\gamma$-ray ($\gamma_{D}$) spectrum for 0~$<~t_{decay}~<$~4~$\mu$s is shown in Fig.~\ref{fig:129sb_fig2}(b), yielding 98, 189, 234, 755 and 1128~keV transitions, as expected from the previous works. The 234 and 755~keV transitions are newly observed (marked with hash). The 234~keV is placed below the $19/2^{+}$ state decaying to $(17/2^{-})$ state, by following the systematics with the 120~keV in $^{127}$ Sb and 108~keV in $^{131}$Sb. The 755~keV on the other hand is assigned to be depopulating the $(15/2^{+}$) to the $11/2^{+}$ state, in accordance with the 852~keV in $^{127}$Sb and the shell model calculations. The required 157~keV from $19/2^{+}$ to $(15/2^{+})$, as seen in the lower odd-A Sb isotopes (442, 332, 247 keV in $^{123-127}$Sb, respectively), is not seen in the delayed spectrum. The half-life fit (one-component) for the decay spectrum upon gating on 189~keV transition yields a value of $T_{1/2}$ = 0.89(3)~$\mu$s (in agreement with the value of 1.1(1) $\mu$s reported in  Ref.~\cite{ge03}), which gives B(E2; $23/2^{+} \rightarrow 19/2^{+}$) = 25.3(8)~e$^{2}$fm$^{4}$ for the $23/2^{+}$ state. Similarly, Fig.~\ref{fig:129sb_fig2}(c) shows the $\gamma_{D}$-ray spectrum for 5~$<~t_{decay}~<$~10~$\mu$s, leading to 732 and 1128~keV transitions. A half-life fit (two-component with one component fixed to 0.89 $\mu$s) for the decay spectrum upon gating on 732~keV transition yields a value of $T_{1/2}$ = 2.3(3)~$\mu$s for the $15/2^{-}$ state (in agreement with the value of 2.2(2)~$\mu$s reported in Ref.~\cite{ge03}). The tracked Doppler corrected $\gamma_{P}$ in coincidence with the sum gate of $\gamma_{D}$'s, namely 98, 189, 234 and 755~keV delayed transitions (0~$<~t_{decay}~<$~4~$\mu$s) is shown in Fig.~\ref{fig:129sb_fig2}(d). This spectrum yields the newly identified prompt 156, 315 and 622~keV $\gamma$-ray transitions. Thus, all these transitions are placed above the $23/2^{+}$ isomer. Since the 156~keV transition is not observed in coincidence with 315 and 622~keV transitions (not shown in this figure), it is assigned depopulating from $(21/2^{+})$ to $23/2^{+}$ state, in accordance with the 161~keV in $^{127}$Sb and 206~keV in $^{131}$Sb. In addition, the $\gamma_{D}$ in coincidence with all these newly observed prompt transitions are studied and all of these yield the 98, 189 and 1128~keV transitions (not shown in this figure). The tracked Doppler corrected $\gamma_{P}$ in coincidence with any $\gamma_{D}$ for 5~$<~t_{decay}~<$~10~$\mu$s, did not result in any new prompt $\gamma$ rays and hence no transitions are placed above the $15/2^{-}$ isomer. No prompt-delayed correlations could be carried out for the  $19/2^{-}$ isomer, as it has a very long half-life of 17.7~min. As the inset of Fig.~\ref{fig:129sb_fig2}(a) shows the coincidence of  530 and 1078~keV transitions and a similar systematics has been observed in the lower odd-A Sb isotopes, these transitions are placed above the $19/2^{-}$ isomer.  

\subsection{\label{sec:130sb}$^{130}$Sb}

\begin{figure}[t]
\includegraphics[width=1.0\columnwidth]{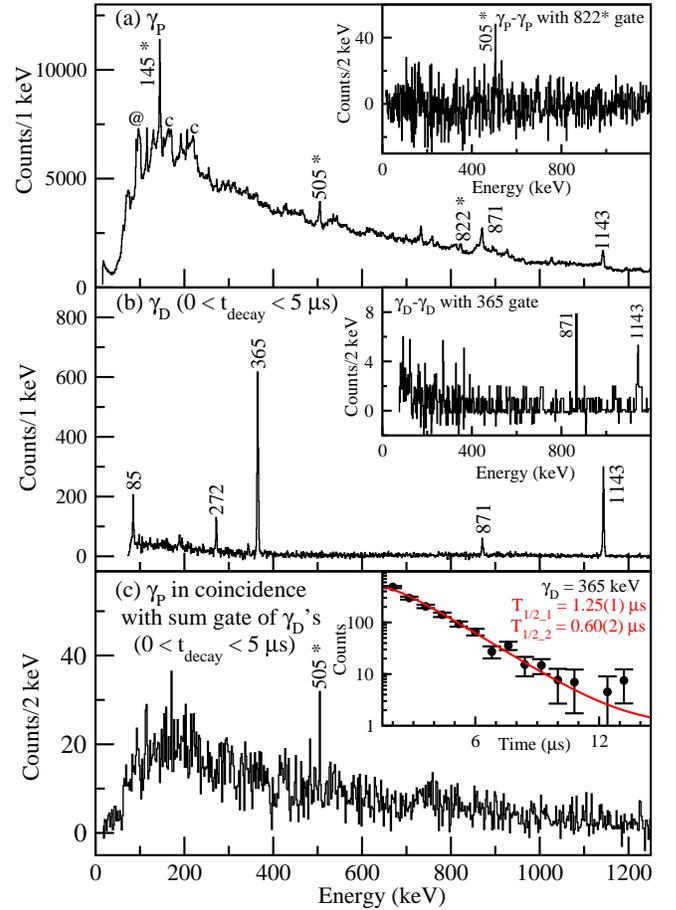}
\caption{\label{fig:130sb_fig2} (Color online) $A-$ and $Z-$gated $\gamma$-ray spectra for $^{130}$Sb: (a) Tracked Doppler corrected prompt singles $\gamma$-ray spectrum ($\gamma_{P}$). The newly identified prompt $\gamma$ rays are marked with an asterisk. The inset shows the tracked Doppler corrected $\gamma_{P}$-$\gamma_{P}$ coincidence spectrum with gate on the newly identified 822~keV prompt $\gamma$-ray. (b) The delayed singles $\gamma$-ray spectrum ($\gamma_{D}$) for 0~$<~t_{decay}~<$~5~$\mu$s. The inset shows the $\gamma_{D}$-$\gamma_{D}$ coincidence spectrum with gate on the delayed 365~keV $\gamma$-ray. (c) The tracked Doppler corrected $\gamma_{P}$ in coincidence with the delayed $\gamma$ rays = 365 and 1143~keV transitions, for 0~$<~t_{decay}~<$~5~$\mu$s. The newly identified prompt transitions are marked with an asterisk. The  inset shows the decay curve along with the two-component fit for the delayed 365~keV transition.}
\end{figure}

\begin{table}[h]
\caption{\label{tab:130sb_tab1}Properties of the transitions assigned to $^{130}$Sb obtained in this work.  The top and bottom panels, separated by a line, are for the prompt and delayed transitions, respectively.}
\resizebox{0.4\textwidth}{!}{
\begin{ruledtabular}
\begin{tabular}{ccccc}

 E$_{\gamma}$ & I$_{\gamma}$ & $J_{i}^{\pi} \rightarrow  J_{f}^{\pi}$ & $E_{i}$ & $E_{f}$\\ \hline\\
 504.9 & 100 & (12$^{+})\rightarrow$ (11$^{+})$  & 2050 & 1545 \\
 821.8 & 48(8) & (13$^{+})\rightarrow$ (12$^{+})$  & 2872 & 2050 \\ \hline\\
 84.7 & 11(10) & 6$^{-}\rightarrow$ 8$^{-}$  & 85 & 0 \\
 272.1 & 23(9) & (10$^{-})\rightarrow$ (9$^{-})$  & 1143 & 871 \\
 365.2 & 113(42) & (11$^{+})\rightarrow$ (10$^{-})$  & 1508 & 1143 \\
 871.1 & 29(7) & (9$^{-})\rightarrow$ 8$^{-}$  & 871 & 0 \\
 1143.4 & 100 & (10$^{-})\rightarrow$ 8$^{-}$  & 1143 & 0 \\ 

\end{tabular}
\end{ruledtabular}}
\end{table}

Previous measurement on the high-spin states using $\gamma$-ray spectroscopy in $^{130}$Sb was reported in Ref.~\cite{ge02}. The level scheme as obtained in the present work is shown in Fig.~\ref{fig:sb_fig1}. Table.~\ref{tab:130sb_tab1} shows the properties of all the transitions assigned in this work.

The $A-$ and $Z-$gated $\gamma$-ray spectra for $^{130}$Sb are shown in Fig.~\ref{fig:130sb_fig2}. The tracked Doppler corrected prompt singles $\gamma$-ray spectrum ($\gamma_{P}$) for $^{130}$Sb is shown in Fig.~\ref{fig:130sb_fig2}(a). The already known 871 and 1143 keV $\gamma$ rays are observed. In addition, three new prompt transitions, namely 145, 505 and 822 keV transitions are observed, which are marked with an asterisk. The inset in Fig.~\ref{fig:130sb_fig2}(a) shows the tracked Doppler corrected prompt $\gamma$-$\gamma$ coincidence spectrum ($\gamma_{P}$-$\gamma_{P}$) with gate on the newly identified 822~keV transition. This shows that the 505 and 822~keV transitions are in coincidence, as shown in the level scheme. But 145~keV is not observed in coincidence and hence not placed in the level scheme. Figure.~\ref{fig:130sb_fig2}(b) shows the delayed $\gamma$-ray ($\gamma_{D}$) singles spectrum with 0 $< t_{decay} <$ 5~$\mu$s. All the known delayed $\gamma$ rays, namely 85, 272, 365, 871 and 1143~keV transitions are observed. The delayed $\gamma$-$\gamma$ ($\gamma_{D}$-$\gamma_{D}$) coincidence spectrum with gate on 365~keV transition is shown in the inset of Fig.~\ref{fig:130sb_fig2}(b). Figure ~\ref{fig:130sb_fig2}(c) shows the tracked Doppler corrected $\gamma_{P}$ in coincidence with the sum gate of $\gamma_{D}$ = 365 and 1143~keV transitions. This yields the newly identified 505~keV transition. As from the inset of Fig.~\ref{fig:130sb_fig2}(a), the 505 and 822~keV transitions are seen in coincidence, these two transitions are placed above the ($11^{+}$) state, following the systematics with the lower even-A Sb isotopes. The inset in Fig.~\ref{fig:130sb_fig2}(c) shows the decay curve for the delayed 365~keV transition. A one-component fit to this transition yielded a value shorter than 1.8(2)~$\mu$s for the ($13^{+}$) state~\cite{ge02}. The discrepancy in the $T_{1/2}$ of ($13^{+}$) state may be due to the time of flight of $\sim$ 2~$\mu$s for the setup used in Ref.~\cite{ge02}, that bias the measurement of the half-life to higher values. Thus, the ($11^{+}$) state might also have a half-life, as observed in the lower even-A Sb isotopes. A two component fit (as shown in the lower inset in Fig.~\ref{fig:130sb_fig2}(c))  was carried out, which yielded a half-life of 1.25(1)~$\mu$s for the (13$^{+}$) state and 0.600(15)~$\mu$s for the ($11^{+}$) state. The B(E2; $13^{+} \rightarrow 11^{+}$) = 105(6)~e$^{2}$fm$^{4}$ was obtained. 

\subsection{\label{sec:131sb}$^{131}$Sb}

\begin{figure}[t]
\includegraphics[width=1.0\columnwidth]{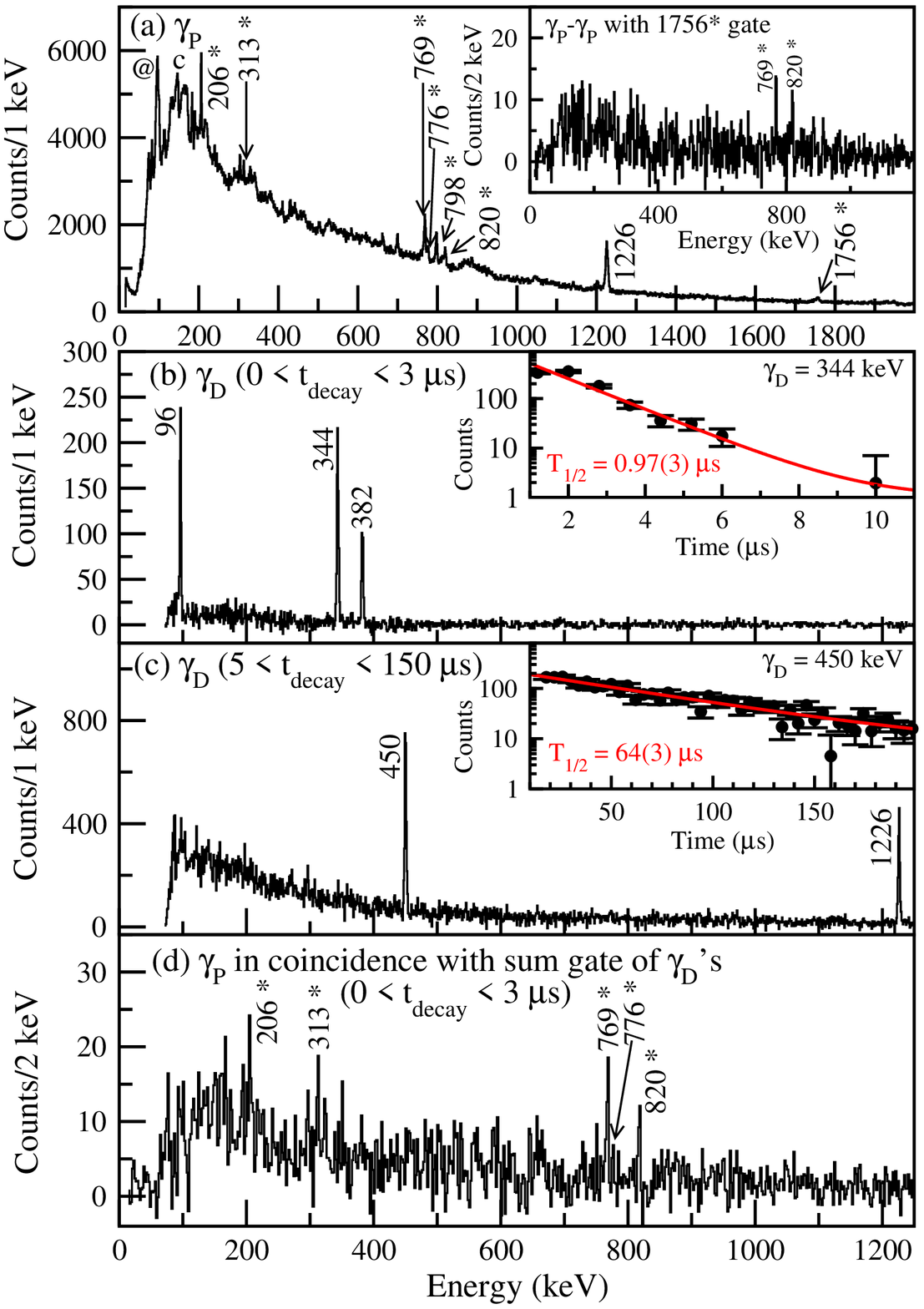}
\caption{\label{fig:131sb_fig2} (Color online) $A-$ and $Z-$gated $\gamma$-ray spectra for $^{131}$Sb: The tracked Doppler corrected prompt singles $\gamma$-ray ($\gamma_{P}$) spectrum with the new $\gamma$-ray transitions marked with asterisk. The inset shows the tracked Doppler corrected prompt $\gamma_{P}$-$\gamma_{P}$ coincidence spectrum with gate on the newly identified 1756 keV transition. (b) and (c) The delayed singles $\gamma$-ray ($\gamma_{D}$) spectra for 0 $< t_{decay} <$ 3 $\mu$s and 5 $< t_{decay} <$ 150 $\mu$s, respectively. The insets in (b) and (c) shows the decay curves along ith the fits for the 344 and 450 keV transitions, respectively. (d) Tracked Doppler corrected $\gamma_{P}$ in coincidence with the sum gate of the $\gamma_{D}$'s, namely 96, 344 and 382 keV delayed transitions (for 0 $< t_{decay} <$ 3 $\mu$s).}
\end{figure}

The high-spin $\gamma$-ray spectroscopy of $^{131}$Sb was previously reported in Refs.~\cite{ge00, sc77}. The level scheme as obtained in the present work is shown in Fig.~\ref{fig:sb_fig1}. Table~\ref{tab:131sb_tab1} shows the properties of all the transitions assigned in this work.

\begin{table}[]
\caption{\label{tab:131sb_tab1}Properties of the transitions assigned to $^{131}$Sb obtained in this work.  The top and bottom panels, separated by a line, are for the prompt and delayed transitions, respectively.}
\resizebox{0.4\textwidth}{!}{
\begin{ruledtabular}
\begin{tabular}{ccccc}

 E$_{\gamma}$ & I$_{\gamma}$ & $J_{i}^{\pi} \rightarrow  J_{f}^{\pi}$ & $E_{i}$ & $E_{f}$\\ \hline\\
 206.1 & 56(4) & (21/2$^{+})\rightarrow$ 23/2$^{+}$  & 2372 & 2166 \\
 313.3 & 16(3) & (25/2$^{+})\rightarrow$ (23/2$^{+})$  & 3461 & 3148 \\
 768.7 & 100 & (25/2$^{+})\rightarrow$ 23/2$^{+}$  & 2935 & 2166 \\
 776.0 & 38(5) & (23/2$^{+})\rightarrow$ (21/2$^{+})$  & 3148 & 2372 \\
 820.0 & 43(4)  & (27/2$^{+})\rightarrow$ (25/2$^{+})$  & 3755 & 2935 \\
 1755.7 & 36(5) & (31/2$^{+})\rightarrow$ (27/2$^{+})$  & 5510 & 3755 \\ \hline\\
 96.3 & 11(8) & 23/2$^{+}\rightarrow$ 19/2$^{+}$  & 2166 & 2070 \\
 343.6 & 30(11) & 19/2$^{+}\rightarrow$ 17/2$^{-}$  & 2070 & 1726 \\
 382.4 & 16(6) & 19/2$^{+}\rightarrow$ 19/2$^{-}$  & 2070 & 1688 \\
 450.0 & 61(22) & 15/2$^{-}\rightarrow$ 11/2$^{+}$  & 1676 & 1226 \\
 1226.2 & 100 & 11/2$^{+}\rightarrow$ 7/2$^{+}$  & 1226 & 0 \\ 

\end{tabular}
\end{ruledtabular}}
\end{table}

\begin{figure*}[]
\includegraphics[width=2.1\columnwidth]{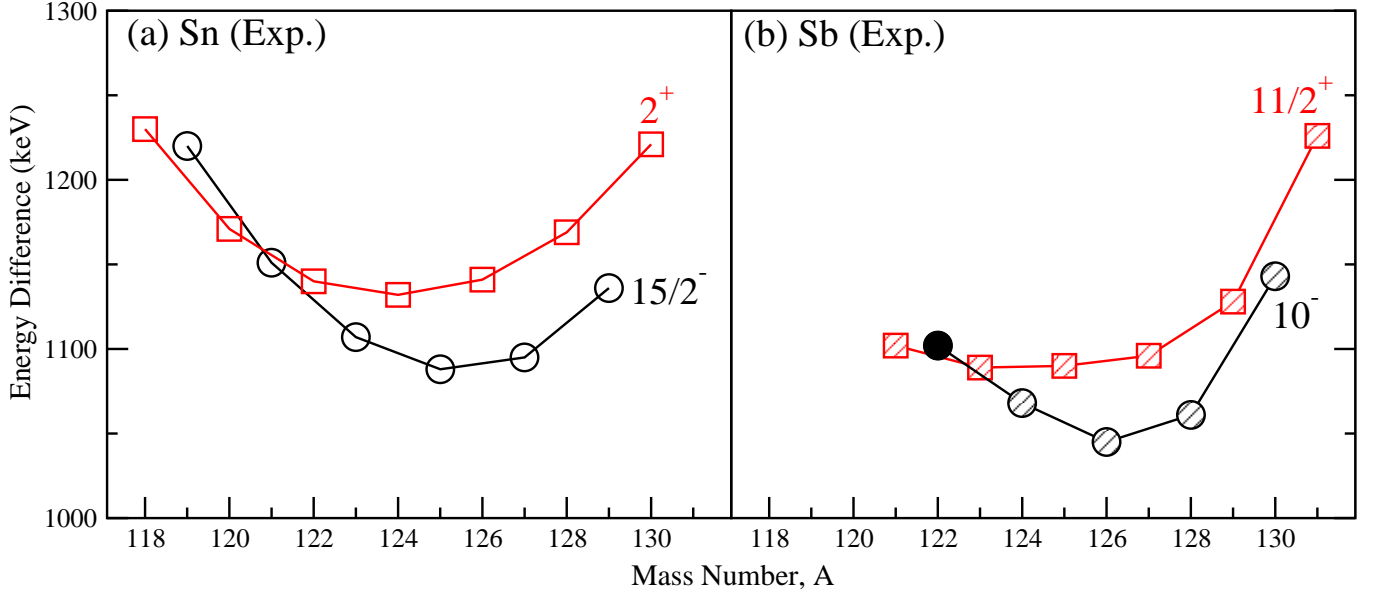}
\caption{\label{fig:dis_fig1} (Color online) Evolution of the experimental energy differences, as a function of mass number, in Sn (open symbols) (a) $E(2^{+})-E(0^{+})$ (red square), $E(15/2^{-})-E(11/2^{-})$ (black circle), and in Sb (shaded symbols) (b) $E(11/2^{+})-E(7/2^{+})$ (red square), $E(10^{-})-E(8^{-})$ (black circle). The filled symbol is the newly observed state from the present experiment. }
\end{figure*}

The $A-$ and $Z-$gated $\gamma$-ray spectra for $^{131}$Sb are shown in Fig.~\ref{fig:131sb_fig2}. The tracked Doppler corrected prompt singles $\gamma$-ray spectrum ($\gamma_{P}$) for $^{131}$Sb is shown in Fig.~\ref{fig:131sb_fig2}(a). The previously known 1226 keV transition is seen in this spectrum. However the other transitions, namely, 96, 344, 382, and 450 keV are not seen in Fig.~\ref{fig:131sb_fig2}(a), as these decay directly from long-lived isomers. In addition, many new prompt $\gamma$-ray transitions, namely 206, 313, 769, 776, 798, 820 and 1756 keV transitions, are identified (marked with an asterisk). The inset shows the tracked Doppler corrected prompt $\gamma_{P}$-$\gamma_{P}$ coincidence spectrum with gate on the newly idientified 1756 keV transition. This spectrum shows that the 769, 820 and 1756 keV transitions are in coincidence. A similar coincidence spectrum is obtained for the 206, 313 and 776 keV transitions (not shown in this figure). However, no coincidences could be observed with a gate on the 798 keV transition, and hence this is not placed in the level scheme. The delayed singles $\gamma$-ray ($\gamma_{D}$) spectrum for 0 $< t_{decay} <$ 3 $\mu$s is shown in Fig.~\ref{fig:131sb_fig2}(b), yielding 96, 344 and 382~keV transitions, as observed in previous measurements. The half-life fit (one-component) for the decay spectrum upon gating on 344~keV transition yields a value of $T_{1/2}$ = 0.97(3)~$\mu$s for the $23/2^{+}$ state (in agreement with the value of 1.1(2)~$\mu$s given in Ref.~\cite{ge00}), yielding B(E2; $23/2^{+} \rightarrow 19/2^{+}$) = 24.6(8)~e$^{2}$fm$^{4}$. Similarly, Fig.~\ref{fig:131sb_fig2}(c) shows the delayed $\gamma$-ray spectrum for 5 $< t_{decay} <$ 150 $\mu$s, leading to 450 and 1226~keV transitions. A half-life fit (using a three component fit with two compnents fixed to 0.97~$\mu$s and 4.3~$\mu$s) for the decay spectrum upon gating on 450~keV transition yields a value of $T_{1/2}$ = 64(3) $\mu$s for the $15/2^{-}$ state (in agreement with 65(5)~$\mu$s quoted by Ref.~\cite{ge00}). The B(E2; $19/2^{-} \rightarrow 15/2^{-}$) = 41(8)~e$^{2}$fm$^{4}$, as given in Ref.~\cite{ju07}. The tracked Doppler corrected $\gamma_{P}$ in coincidence with the sum gate of the $\gamma_{D}$'s, namely 96, 344 and 382~keV delayed transitions (for 0 $< t_{decay} <$ 3~$\mu$s) is shown in Fig.~\ref{fig:131sb_fig2}(d). This spectrum yields almost all the newly identified prompt $\gamma$-ray transitions, that are placed above the $23/2^{+}$ isomer. In addition, the $\gamma_{D}$ in coincidence with all the newly observed prompt transitions are studied and all of these yield the 96, 344 and 382 keV transitions (not shown in this figure). However, the tracked Doppler corrected $\gamma_{P}$ in coincidence with any $\gamma_{D}$ for 5 $< t_{decay} <$ 150 $\mu$s, did not result in any new prompt $\gamma$ rays and hence no transitions are placed above the $15/2^{-}$ isomer. Also, no transitions could be placed above the $19/2^{-}$ isomer.

\section{\label{sec:Dis}Discussion}

\begin{figure*}[]
\includegraphics[width=2.1\columnwidth]{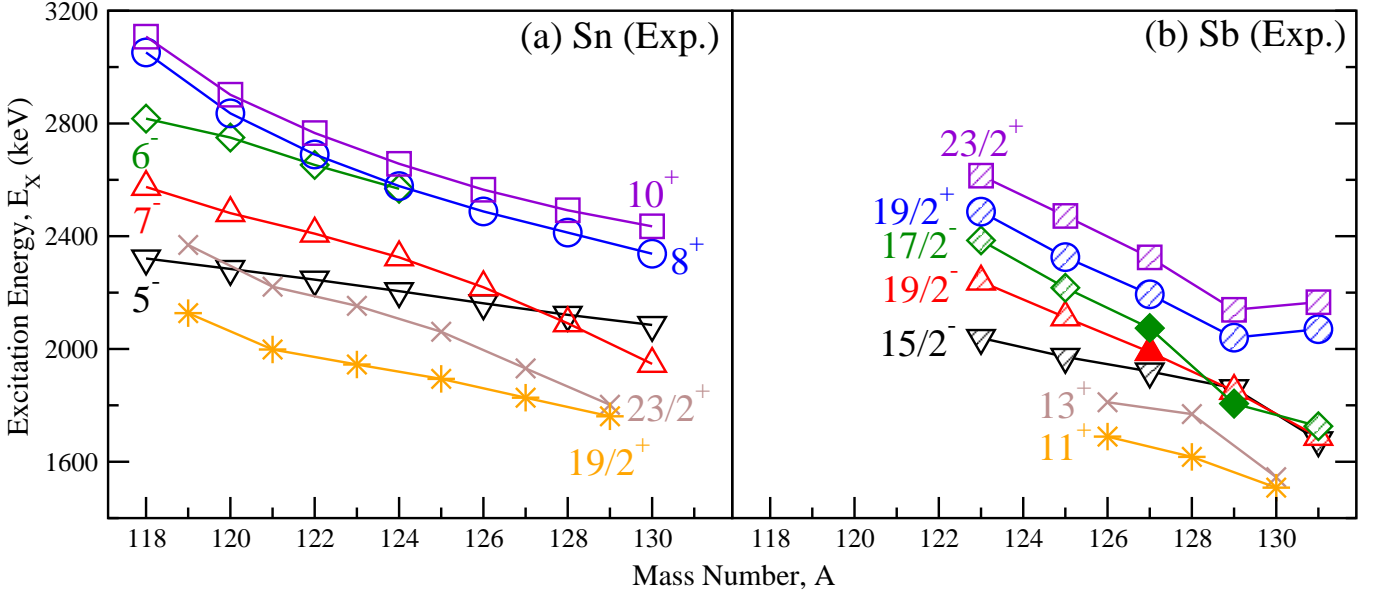}
\caption{\label{fig:dis_fig2} (Color online) Evolution of the experimental (a) $5^{-}$ (black triangle down), $6^{-}$ (green diamond), $7^{-}$ (red triangle up), $8^{+}$ (blue circle), $10^{+}$ (violet square), $23/2^{+}$ (brown x), and $19/2^{+}$ (orange star) states in $^{118-130}$Sn (open symbols), and (b) $15/2^{-}$ (black triangle down), $17/2^{-}$ (green diamond), $19/2^{-}$ (red triangle up), $19/2^{+}$ (blue circle), $23/2^{+}$ (violet square), $13^{+}$ (brown x), and $11^{+}$ (orange star) states in $^{123-131}$Sb (shaded symbols) with mass number, A. The filled symbols are the newly observed states from the present experiment.}
\end{figure*} 

\begin{figure*}[]
\includegraphics[width=2.1\columnwidth]{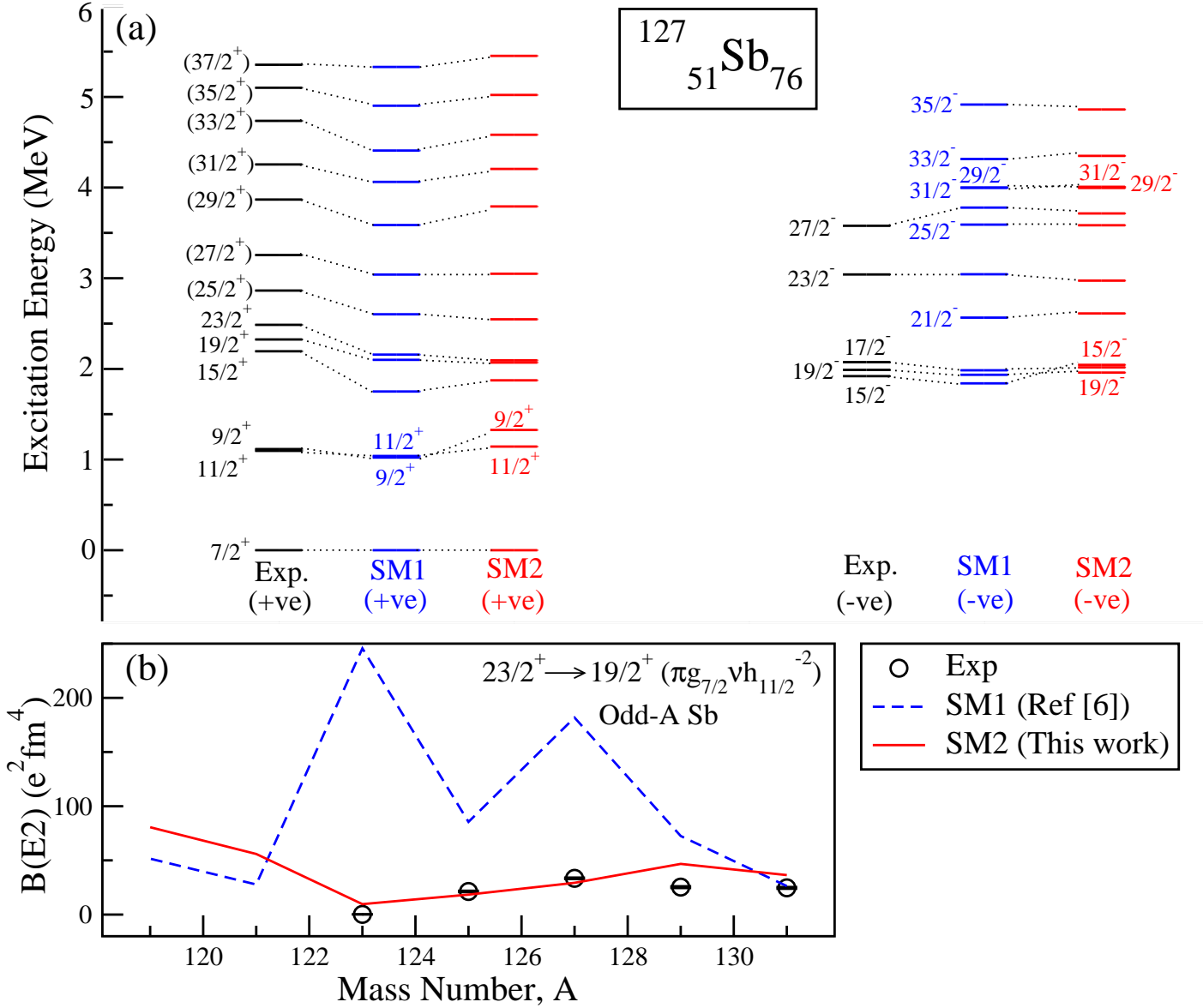}
\caption{\label{fig:dis_fig3} (Color online) (a) Comparison of the experimental level schemes with that obtained from shell model calculations using the interaction in Ref.~\cite{re16} (SM1 blue) and the modified interaction in the present work (SM2 red) for both positive and negative-parity states in $^{127}$Sb. The different states with the same spin have been joined by dotted lines to show the agreement. (b) The experimental and calculated B(E2) values (black square) for the $23/2^{+} \rightarrow 19/2^{+}$ in odd-A $^{123-131}$Sb (see text). }
\end{figure*}

\begin{figure*}[]
\includegraphics[width=2.1\columnwidth]{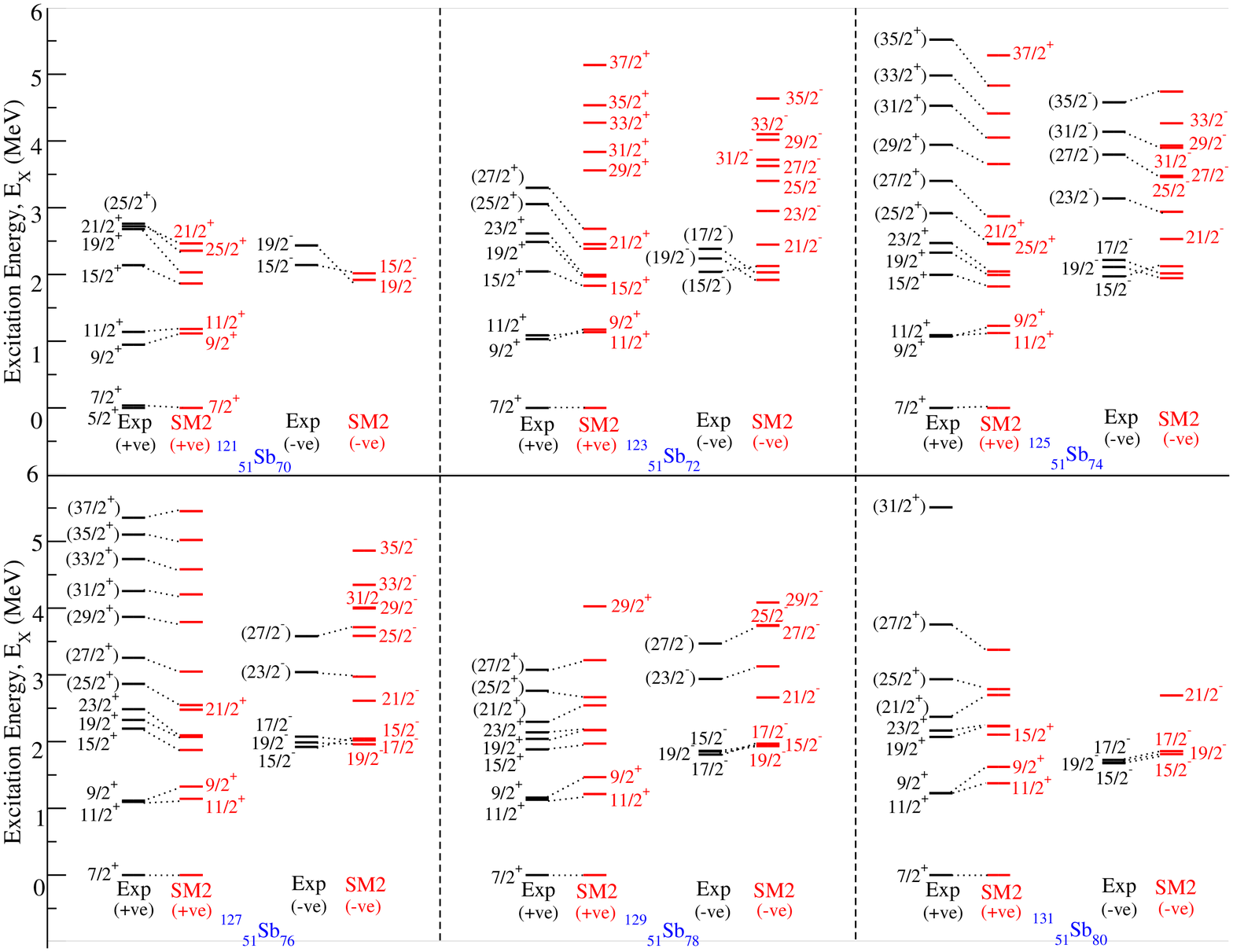}
\caption{\label{fig:dis_fig4} (Color online) Comparison of experimental (Exp.) (shown in black)~\cite{ju07, wa09, wa09epja, ge03, ge00} and theoretical calculations (from shell model (SM2)) (shown in red) level schemes for odd-A $^{121-131}$Sb isotopes for both positive (+ve) and negative (-ve) parities. The experimental and shell model states of the same spin-parity have been joined with dotted lines to show the agreement.}
\end{figure*}

\begin{figure*}[]
\includegraphics[width=2.1\columnwidth]{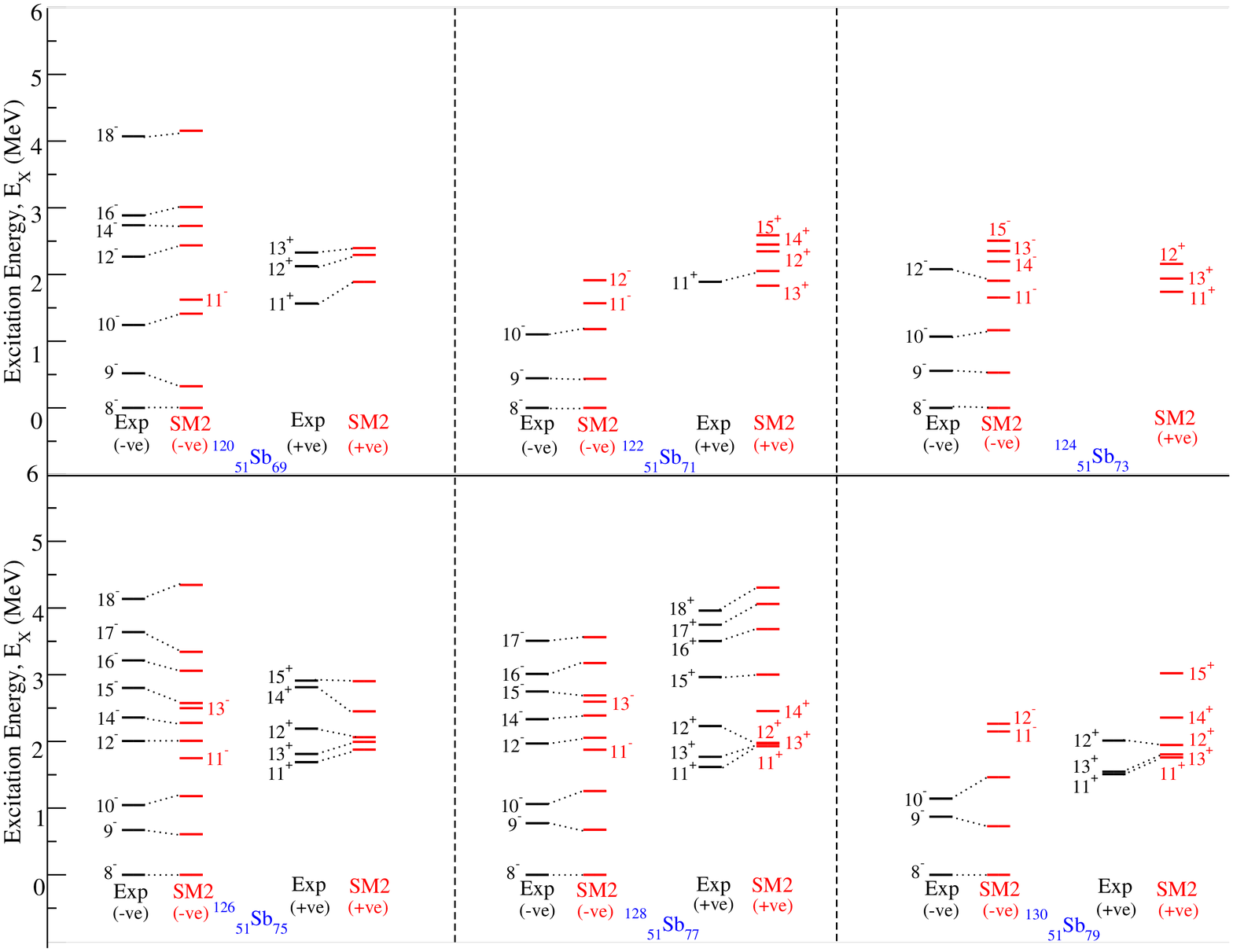}
\caption{\label{fig:dis_fig5} (Color online) Comparison of experimental (Exp.) (shown in black)~\cite{li14, gu77, la88, re16, ge02} and theoretical calculations (from shell model (SM2)) (shown in red) level schemes for even-A $^{122-130}$Sb isotopes for both positive (+ve) and negative (-ve) parities. The experimental and shell model states of the same spin-parity are joined with dotted lines to show the agreement.}
\end{figure*}

\begin{figure*}[]
\includegraphics[width=2.1\columnwidth]{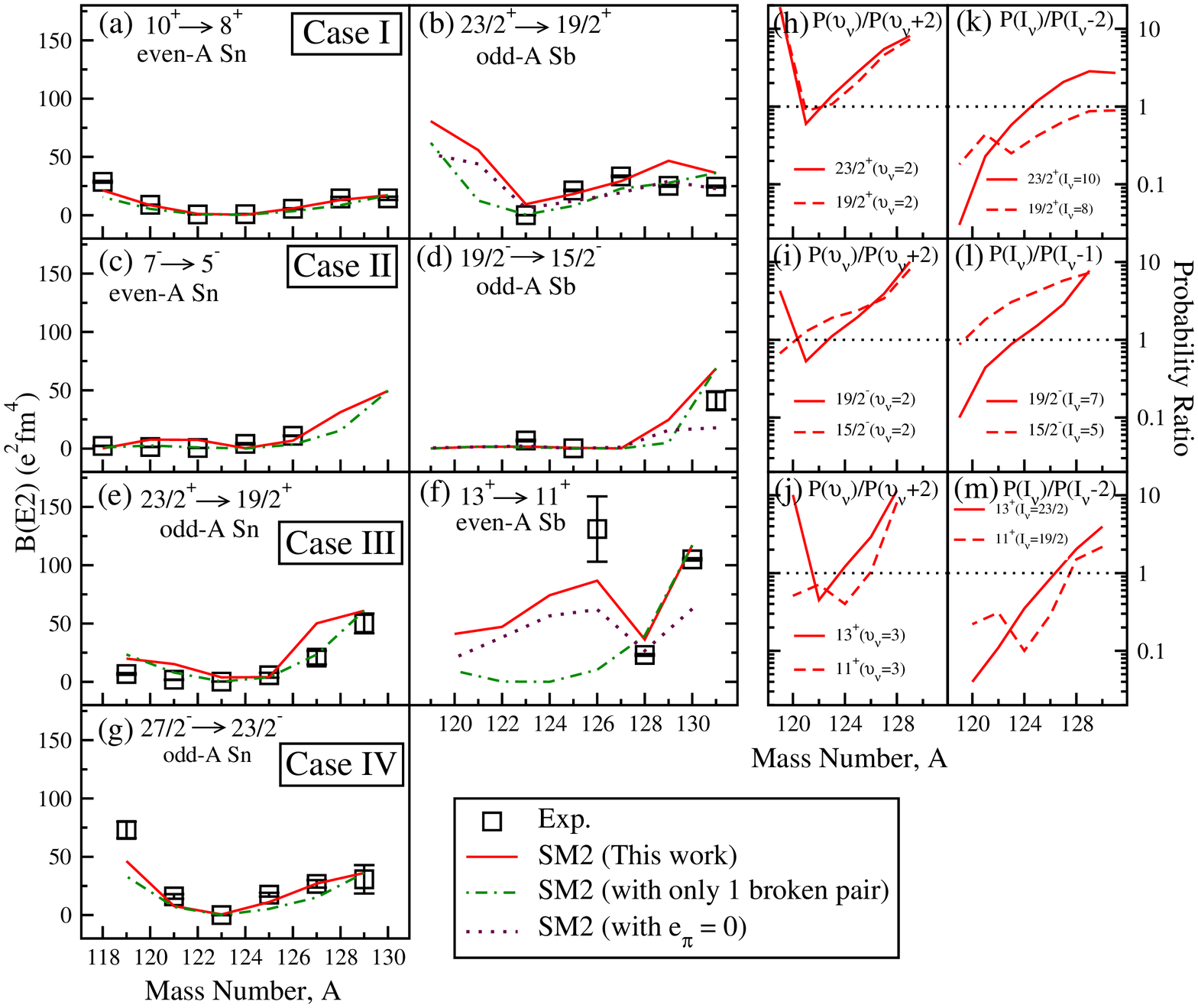}
\caption{\label{fig:dis_fig6}(Color online) The experimental B(E2) values (black square) for the (a)$10^{+} \rightarrow 8^{+}$ in even-A $^{118-130}$Sn~\cite{fo81, br92, is16}, (b) $23/2^{+} \rightarrow 19/2^{+}$ in odd-A $^{123-131}$Sb (see text) (c) $7^{-} \rightarrow 5^{-}$ in even-A $^{118-126}$Sn~\cite{is16}, (d) $19/2^{-} \rightarrow 15/2^{-}$ transitions in odd-A $^{123-131}$Sb (see text), (e) $23/2^{+} \rightarrow 19/2^{+}$ in odd-A $^{119-129}$Sn~\cite{lo08, is16}, (f) $13^{+}$ and $11^{+}$ states in the even-A $^{126-130}$Sb (see text), and (g) $27/2^{-} \rightarrow 23/2^{-}$ in odd-A $^{119-129}$Sn isotopes~\cite{ma94, lo08, is16}. The shell model calculations using the full interaction SM2 (red solid line), SM2 restricted to only one broken neutron pair (green dash-dotted line) and SM2 with $e_{\pi} = 0$ (purple dotted line) are also shown. The probability ratios for the (h-j) neutron seniority mixing ($(P(\upsilon_{\nu})/P(\upsilon_{\nu}+2)$) and (k-m) neutron-angular momentum mixing ($(P(I_{\nu})/P(I_{\nu}-2)$) and ($(P(I_{\nu})/P(I_{\nu}-1)$) are also shown.}
\end{figure*}

A systematic study of both odd-A and even-A $^{122-131}$Sb isotopes was carried out in this work. The observed excited states in $^{122-131}$Sb (Z = 51) isotopes have a close correspondence with those in $^{121-130}$Sn (Z = 50) isotopes.  This can be evidenced from the similarities in the energy differences of the low-lying states in the even-A/odd-A Sn {\it i.e.} $E(2^{+})-E(0^{+}) \sim E(15/2^{-})-E(11/2^{-})$ and odd-A/even-A Sb {\it i.e.} $E(11/2^{+})-E(7/2^{+}) \sim E(10^{-})-E(8^{-})$, respectively. This is depicted in Fig.~\ref{fig:dis_fig1}. Similar correspondence for selected high-spin states is presented in Fig.~\ref{fig:dis_fig2}. These similarities are due to the fact that the Sb isotopes have a single valence proton particle in the $g_{7/2}$ orbital in addition to neutrons in the corresponding Sn isotopes. 

A better understanding of the aforementioned correspondence for the high-spin states in Sb and Sn isotopes was achieved by performing shell model calculations, with the interaction used in Ref.~\cite{re16} (denoted by SM1). The model space was constituted of a restricted single-particle space consisting of active particles, the neutrons in ($\nu$) $d_{3/2}$, $s_{1/2}$, $h_{11/2}$ orbits, and a proton in ($\pi$) $g_{7/2}$ orbit near the Fermi surface. This interaction was derived from the original jj55pn interaction~\cite{hj95}, where the {\it diagonal} Two-Body Matrix Elements (TBMEs) were adjusted to account for the missing correlations in the restricted model space. The modified~(original) multiplets, used in Ref.~\cite{re16} were:\\
(i) $\langle d_{3/2}^{2};I \arrowvert \hat{\mathcal{H}}\arrowvert d_{3/2}^{2};I\rangle$ = -0.6048~(-0.3024), and 0.22104~(0.14814) for $I = 0$ and 2, respectively;\\
(ii) $\langle s_{1/2}^{2};I \arrowvert \hat{\mathcal{H}}\arrowvert s_{1/2}^{2};I\rangle$ = -0.959904~(-0.59994) for $I = 0$;\\
(iii) $\langle h_{11/2}^{2};I \arrowvert \hat{\mathcal{H}}\arrowvert h_{11/2}^{2};I\rangle$ = -1.88544~(-0.94545), -0.88533~(-0.435330), 0.20683~(0.10683), 0.27145~(0.17145), and 0.33148~(0.23148) for $I = 0, 2, 6, 8$ and 10, respectively;\\
(iv) $\langle d_{3/2}h_{11/2};I \arrowvert \hat{\mathcal{H}}\arrowvert d_{3/2}h_{11/2};I\rangle$ = -0.08335~(0.016650) for $I = 4$; and\\
(v) $\langle d_{3/2}s_{1/2};I \arrowvert \hat{\mathcal{H}}\arrowvert d_{3/2}s_{1/2};I\rangle$ = 0.05607~(-0.06399), for $I = 2$;\\ 
where $\hat{\mathcal{H}}$ represents the Hamiltonian.
Effective charges of $e_{\nu}$ = 0.9 and $e_{\pi}$ = 1.8 were chosen to reproduce the known $B(E2)$ values in $^{130}$Sn $(B(E2);~10^{+} \rightarrow 8^{+})$ and $^{134}$Te $(B(E2);~6^{+} \rightarrow 4^{+})$ isotopes. The calculations were performed using the shell model code, NATHAN~\cite{ca99}. Using SM1, the excitation energies could be well reproduced, while the binding energies and $B(E2)$ transition probabilities were not properly reproduced. This is illustrated for $^{127}$Sb for the level energies and for the $B(E2;~23/2^{+} \rightarrow 19/2^{+})$ in odd-A Sb isotopes in Fig.~\ref{fig:dis_fig3}. Similar inconsistencies were observed for the other isomeric transitions (not shown in the figure). 

Therefore, to improve the agreement of $B(E2)$ values, the following parameters of the shell model interaction were modified (denoted by SM2):\\
(i) the monopole part of $\langle h_{11/2}^{2}\arrowvert \hat{\mathcal{H}}\arrowvert h_{11/2}^{2}\rangle$ was reduced by 185 keV, which led to an improved reproduction of the binding energies and the $(B(E2);~10^{+} \rightarrow 8^{+})$ for the even-A Sn isotopes;\\
(ii) the monopole part of  $\langle d_{3/2}^{2}\arrowvert \hat{\mathcal{H}}\arrowvert d_{3/2}^{2}\rangle$ was reduced by 450 keV, which led to a better reproduction of the $(B(E2);~23/2^{+} \rightarrow 19/2^{+})$ in odd-A Sb isotopes;\\
(iii) the pairing term, $\langle d_{3/2}^{2}\arrowvert \hat{\mathcal{H}}\arrowvert h_{11/2}^{2}\rangle$, was increased by 260 keV, which led to the agreement of all the $B(E2)$ values, except those for the odd-A Sb isotopes. Also, all the excitation energies became higher by 2 MeV;\\
(iv) the pairing term, $\langle h_{11/2}^{2}\arrowvert \hat{\mathcal{H}}\arrowvert h_{11/2}^{2}\rangle$, was increased by 140 keV, to compensate for this increase in excitation energies as well as seeing to it that the $B(E2)$'s of all the Sn and Sb do not change drastically;\\
(v) the $\langle \pi g_{7/2} \nu h_{11/2};I \arrowvert \hat{\mathcal{H}}\arrowvert \pi g_{7/2} \nu h_{11/2};I\rangle$ ($I = 8$) was reduced by 400 keV, which reproduced the $B(E2)$'s for the odd-A Sb as well; and \\
(vi) the monopole part of $\langle d_{3/2}h_{11/2}\arrowvert \hat{\mathcal{H}}\arrowvert d_{3/2}h_{11/2}\rangle$ was increased by 30 keV to achieve a better reproduction of $B(E2)$ in odd-A Sn isotopes. \\
These modifications led to the reproduction of the binding energies (within $\sim$~2 MeV), excitation energies (within $\sim$~500 keV) and the $B(E2)$ transition strengths (within $\sim$~50 $e^{2}fm^{4}$) for the isotopic chains of $^{119-130}$Sn and $^{121-131}$Sb. The energies of the excited states for $^{127}$Sb and the improved $B(E2)$ values for the odd-A Sb isotopes are shown in Fig.~\ref{fig:dis_fig3}. The reason behind the large discrepancy in the $B(E2)$'s between SM1 and SM2 can be understood based on the difference in the wavefunctions obtained in both calculations. The wavefunctions from SM2 are much more fragmented configuration wise than those in SM1, leading to the dramatic reduction in the $B(E2)$ values.

\subsection{\label{sec:Energy} Excitation energies}

A comparison of the experimental (black) and calculated (red) level schemes (using SM2) for both the positive (+ve) and negative (-ve) parities in odd-A $^{123-131}$Sb isotopes is shown in Fig.~\ref{fig:dis_fig4}. The data for $^{121}$Sb is taken from Ref.~\cite{wa09} and is also shown in this figure for comparison. The calculated order for the $9/2^{+}$ and $11/2^{+}$ states are inverted only for the odd-A $^{123-125}$Sb isotopes. A mismatch in the ordering of the $15/2^{-}$, $17/2^{-}$ and $19/2^{-}$ states are observed in $^{121-129}$Sb isotopes, as these states are very close in energy. Similarly, Fig.~\ref{fig:dis_fig5} shows the comparison of the experimental (black) and calculated (red) level schemes for both the positive (+ve) and negative (-ve) parities in even-A $^{122-130}$Sb isotopes. The data for $^{120}$Sb is taken from Ref.~\cite{li14} and is also shown in this figure for comparison. This figure shows that the experimental energies are in reasonable agreement with the theoretical calculations.

 Isomeric $15/2^{-}$, $19/2^{-}$, and $23/2^{+}$ were previously observed in odd-A Sb isotopes. These were interpreted as the odd $g_{7/2}$ proton coupled to the known isomeric $5^{-}$, $7^{-}$, and $10^{+}$ states in the corresponding even-(A-1) Sn isotopes, respectively~\cite{ap74}. These states along with the $6^{-}$ and $8^{+}$ states for even-A Sn and the $17/2^{-}$ and $19/2^{+}$ states for odd-A Sb are shown in Fig.~\ref{fig:dis_fig2}(a) and (b), respectively. In addition, the evolution of $19/2^{+}$ and $23/2^{+}$ states in odd-A Sn; and $11^{+}$ and $13^{+}$ states in even-A Sb are shown in these plots. The newly identified $(17/2^{-}$) and $(19/2^{-}$) states in $^{127}$Sb; and the $(17/2^{-}$) state in $^{129}$Sb (from the current experimental analysis) fit in the systematics. Comparison with the Sn isotopes shows that the energies of the positive-parity states in Sb follow a similar pattern to that of the Sn isotopes, with a slight dip in the case of $^{129}$Sb. Also, the crossing of the $7^{-}$ and $5^{-}$ observed at $^{128}$Sn is seen in the equivalent states in $^{129}$Sb. In addition, the newly identified $11^{+}$ and $13^{+}$ states in even-A Sb isotopes follow a similar pattern to the corresponding $19/2^{+}$ and $23/2^{+}$ states in odd-A Sn isotopes.

\subsection{\label{sec:B(E2)} $B(E2)$ Transition probabilities}

The experimental $B(E2)$ values along with SM2 shell model calculations are shown in Fig.~\ref{fig:dis_fig6}.\\
{\bf Case I}: (a) $10^{+} \rightarrow 8^{+}$ in even-A $^{118-130}$Sn~\cite{fo81, br92, is16}, (b) $23/2^{+} \rightarrow 19/2^{+}$ in odd-A $^{123-131}$Sb (see text),\\ 
{\bf Case II}: (c) $7^{-} \rightarrow 5^{-}$ in even-A $^{118-126}$Sn~\cite{is16}, (d) $19/2^{-} \rightarrow 15/2^{-}$ in odd-A $^{123-131}$Sb (see text), \\
{\bf Case III}: (e) $23/2^{+} \rightarrow 19/2^{+}$ in odd-A $^{119-129}$Sn~\cite{lo08, is16}, (f) $13^{+} \rightarrow 11^{+}$ states in the even-A $^{126-130}$Sb (see text), and\\ 
{\bf Case IV}: (g) $27/2^{-} \rightarrow 23/2^{-}$ in odd-A $^{119-129}$Sn isotopes~\cite{ma94, lo08, is16}.\\
The full calculations are shown by red solid lines, those restricted to one broken pair are shown by dash-dotted green lines, and the full calculations with proton effective charge, $e_{\pi}$ = 0, are shown by dotted violet lines. The seniority mixing, represented by the ratio of probabilities of lowest, natural ($\upsilon_{\nu}$) and higher ($\upsilon_{\nu} + 2$) neutron seniorities $(P(\upsilon_{\nu})/P(\upsilon_{\nu}+2))$, are shown in Fig.~\ref{fig:dis_fig6} (h-j) for Cases I-III, respectively. The dotted black line with $P(\upsilon_{\nu})/P(\upsilon_{\nu}+2)$ = 1 represents that the mixing of seniorities, $\upsilon_{\nu}$ and $\upsilon_{\nu} + 2$, are equal. Similarly, Fig.~\ref{fig:dis_fig6} (k-m) for Cases I-III, respectively, denote the neutron angular momentum mixing, represented by the ratio of probabilities of highest ($I_{\nu}$) and lower ($I_{\nu} - 2$ or $I_{\nu} - 1$) neutron angular momenta $(P(I_{\nu})/P(I_{\nu}-2)$ or $P(I_{\nu})/P(I_{\nu}-1))$. The dotted black line, similar to the case of neutron seniorities, denotes that the mixing of $I_{\nu}$ and $I_{\nu} - 2$ or $I_{\nu} - 1$ are equal.

\begin{figure*}[]
\includegraphics[width=2.1\columnwidth]{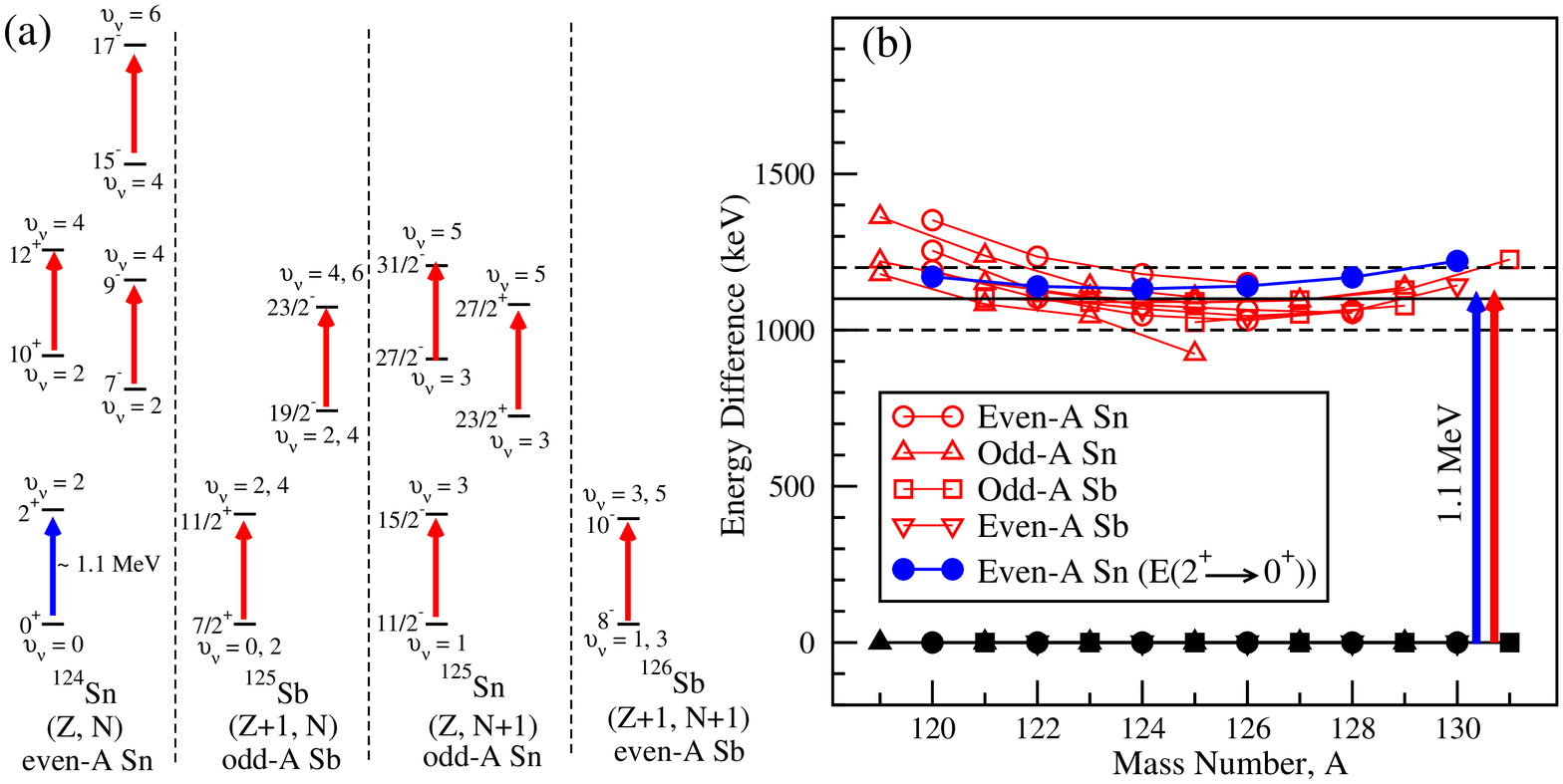}
\caption{\label{fig:dis_fig8} (Color online) (a) The states with different neutron seniorities at different excitation energies in even-A Sn ($Z, N$) ($^{124}$Sn), odd-A Sb ($Z+1, N$) ($^{125}$Sb), odd-A Sn ($Z, N+1$) ($^{125}$Sn), and even-A Sb ($Z+1, N+1$) ($^{126}$Sb) isotopes. The blue and red arrows represent the energy required to break the first pair of neutrons from seniority $\upsilon_{\nu} = n$ to $\upsilon_{\nu} = n+2$ ($n = 0, ..., 4$) in even-A Sn ($E(2^{+} \rightarrow 0^{+})$) and other isotopes, respectively. (b) Plot of the energy differences between states with neutron seniority $\upsilon_{\nu} = n$ and $\upsilon_{\nu} = n+2$ ($n = 0, ..., 4$), as shown in (a), for odd-A and even-A  $^{119-130}$Sn and $^{121-131}$Sb nuclei.}
\end{figure*}

The present calculations agree well with the experimental values and the experimental trends are reproduced. In general, the shape of the $B(E2)$ curves for Sn follow a parabolic behaviour, as expected for seniority scheme. The full calculation and the one restricted to one broken pair, in Sn, are very similar and also the calculated neutron seniority $\upsilon_{\nu}$ = 2, 3 dominates for odd-A, even-A Sn, respectively. In addition, the behaviour of $B(E2)$ for odd-A Sb isotopes is similar to that of the corresponding Sn, except the even-A Sb. The proton charge ($e_{\pi}$) does not have a considerable impact on the $B(E2)$, except in even-A Sb. A detailed discussion of the four different cases, mentioned above, is given below:\\

{\bf Case I:} From shell model calculations, it is seen that the dominant configuration of the $10^{+}$ and $8^{+}$ states in even-A Sn is $\nu h_{11/2}^{-2}$, while that of the $23/2^{+}$ and $19/2^{+}$ states in odd-A Sb is $\pi g_{7/2} \nu h_{11/2}^{-2}$. The nature of the curves in Fig.~\ref{fig:dis_fig6} (a) and (b) is not completely identical. This is evident from the differences in the full calculations and that restricted to one broken pair in Sb. For a microscopic understanding behind this difference, a comparison of the neutron seniority for the $23/2^{+}$ and $19/2^{+}$ states was carried out. Figure.~\ref{fig:dis_fig6} (h) shows that the lowest neutron seniority $\upsilon_{\nu} = 2$ is mixed with $\upsilon_{\nu} = 4$ for both states. The mixing increases with the increasing number of valence neutron holes. However, for the lowest A, the number of available neutron pairs is reduced in the restricted model space, which leads to a sudden drastic decrease in seniority mixing.
In addition to mixing of neutron seniorities, neutron angular momentum  mixing is also observed, as shown in Fig.~\ref{fig:dis_fig6} (k). The angular momentum mixing also increases with increase in the number of valence neutron holes. A comparison of neutron seniority and neutron angular momentum mixing plots shows that the angular momentum mixing is the dominant one. \\

{\bf Case II:} Shell model calculations show that the dominant configuration of the $7^{-}$ and $5^{-}$ states in even-A Sn is $\nu h_{11/2}^{-1}d_{3/2}^{-1}$, while that of the $19/2^{-}$ and $15/2^{-}$ states in odd-A Sb is $\pi g_{7/2} \nu h_{11/2}^{-1}d_{3/2}^{-1}$. The nature of the curves in Fig.~\ref{fig:dis_fig6} (c) and (d) is similar. A comparison of the neutron seniority for the $19/2^{-}$ and $15/2^{-}$ states was carried out. Figure.~\ref{fig:dis_fig6} (i) shows that the lowest neutron seniority $\upsilon_{\nu} = 2$ is mixed with the $\upsilon_{\nu} = 4$ for both the states, similar to {\bf Case I}. In addition to mixing of neutron seniorities, neutron angular momentum mixing is also observed, as shown in Fig.~\ref{fig:dis_fig6} (l), which is of similar order as the seniority mixing. The angular momentum mixing also increases with increase in the number of valence neutron holes.\\

{\bf Case III:} The shell model calculations show that the dominant configuration of the $23/2^{+}$ and $19/2^{+}$ states in odd-A Sn is $\nu h_{11/2}^{-2}d_{3/2}^{-1}$, while that of the $13^{+}$ and $11^{+}$ states in even-A Sb is $\pi g_{7/2} \nu h_{11/2}^{-2}d_{3/2}^{-1}$. The nature of the curves in Fig.~\ref{fig:dis_fig6} (e) and (f) is significantly different. This is also evident from the differences in the full calculations and that restricted to one broken pair in Sb. The calculations with proton charge restricted to 0 show that the neutrons have a considerable impact on the $B(E2)$ in even-A Sb. For a microscopic understanding behind this difference, a comparison of the neutron seniority for the $23/2^{+}$ and $19/2^{+}$ states was carried out. Figure.~\ref{fig:dis_fig6} (j) shows that the lowest neutron seniority $\upsilon_{\nu} = 3$ is mixed with the $\upsilon_{\nu} = 5$ for both states. The mixing increases with the increase in the number of valence neutron holes. In addition to mixing of neutron seniorities, neutron angular momentum mixing is also observed, as shown in Fig.~\ref{fig:dis_fig6} (m). The angular momentum mixing also increases with increase in the number of valence neutron holes. A comparison of neutron seniority and neutron angular momentum mixing plots show that the angular momentum mixing is the dominant one, just like in the {\bf Case I}. It can be noted that around $A = 126$, as compared to the earlier two cases, both the seniority and angular momentum mixing are significant; the higher seniority ($\upsilon_{\nu} + 2$) and the lower angular momentum ($I_{\nu} - 2$) dominate the $11^{+}$ and $13^{+}$ states.\\

{\bf Case IV:} The shell model calculations show that the dominant configuration of the $27/2^{-}$ and $23/2^{-}$ states in odd-A Sn is $\nu h_{11/2}^{-3}$. The corresponding isomer in even-A Sb with configuration $\pi g_{7/2}\nu h_{11/2}^{-3}$ is expected at $I^{\pi} = 16^{-}$. In the present work, the $16^{-}$ states in $^{126, 128}$Sb decay by $M1$ transitions to $15^{-}$ states, and hence no isomer was observed. In $^{120}$Sb, in contrast, the $16^{-}$ state decays via a 148 keV transition to the $14^{-}$ state and has a half-life $T_{1/2}~=~14(3)$ ns~\cite{li14}. The nature of the curve in Fig.~\ref{fig:dis_fig6} (g) is a parabola, similar to Fig.~\ref{fig:dis_fig6} (a). The present calculations show that the dominant seniority for these states is $\upsilon_{\nu}$ = 3, which in fact is in agreement with the calculations assuming only one broken pair.\\

\subsection{\label{sec:Energy} Neutron pair breaking energies}

A striking feature of the constancy of the energy differences, where the increase in the number of broken neutron pairs is involved, was observed and is shown in Fig.~\ref{fig:dis_fig8}. Figure~\ref{fig:dis_fig8}(a) shows particular states in even-A Sn ($Z, N$) ($^{124}$Sn), and odd-A Sn ($Z, N+1$) ($^{125}$Sn) isotopes, which involve the breaking of a pair of neutrons, leading to increase in seniority from $\upsilon_{\nu} = n$ to $\upsilon_{\nu} = n + 2$ , where $n = 0, ..., 4$ at different excitation energies. Similar states in odd-A Sb ($Z+1, N$) ($^{125}$Sb) and even-A Sb ($Z+1, N+1$) ($^{126}$Sb) isotopes are also shown in the same figure. All the arrows represent the energy difference quantum with magnitude of $\sim$ 1.1 MeV. The blue arrow represents breaking of first pair of neutrons in even-A Sn ($E(2^{+} \rightarrow 0^{+}$)). The red arrows are for all the other transitions. A plot of the energy differences in $^{119-130}$Sn and $^{121-131}$Sb isotopes is shown in Fig.~\ref{fig:dis_fig8}(b). A subset of this figure, including only the lower spins, was shown earlier in Fig.~\ref{fig:dis_fig1}. This figure shows that the average energy for the breaking of the first and second pair of neutrons is $\sim$ 1.1 MeV, and this is constant (with a deviation of $\sim$ 100 keV) for a wide range of mass numbers, irrespective of the excitation energy and mixing of neutron seniorities ($\upsilon_{\nu}$) in the case of Sn and Sb. In addition, it follows the behaviour of even-A Sn ($E(2^{+} \rightarrow 0^{+})$) isotopes.

\section{\label{sec:SumCon}Summary and Conclusions}

The neutron-rich $^{122-131}$Sb isotopes were produced as fission fragments in the reaction $^{9}$Be ($^{238}$U, f) with 6.2 MeV/u beam energy. The isomers already known in the odd-A $^{123-131}$Sb isotopes were confirmed and a number of new prompt and delayed transitions were identified for all these isotopes. New isomers, prompt and delayed $\gamma$ rays were identified in the even-A $^{122-130}$Sb isotopes. These results could be achieved using the unique combination of AGATA, VAMOS++ and EXOGAM detectors, which was used for the prompt-delayed spectroscopy of fission fragments. A good agreement was achieved between the experiment and the results of the shell-model calculations in the restricted model space using an optimised interaction. Also the level schemes were in good agreement with that of the corresponding states in the Sn isotopes. The presence of a single valence proton particle in the $g_{7/2}$ orbital leads to a neutron seniority mixing and neutron angular momentum mixing in the Sb isotopes for all the states. The angular momentum mixing, in general, was found to be significantly stronger than the seniority mixing. This work shows that further experimental work is required to search for new isomers in the even-A $^{122,124}$Sb isotopes, which could not be observed in this work either due to low statistics, too long half-life or too short half-life. This will give more insight into the nature of the $B(E2)$ values, and hence the mixing of seniorities and angular momenta due to the $\nu\pi$ interaction.

\section{\label{sec:Ack}Acknowledgements}

The authors would like to thank the AGATA Collaboration for the availability of the AGATA $\gamma$-ray tracking array at GANIL. We acknowledge the important technical contributions of GANIL accelerator staff.  We thank C. Schmitt and A. O. Macchiavelli for help in various aspects of data collection. We also thank P. Van Isacker for his valuable discussions on the theoretical interpretation and careful reading of the manuscript. PB and AM acknowledge support from the Polish National Science Centre (NCN) under Contract No. 2016/22/M/ST2/00269 and the French LEA COPIGAL project. SBi, RB, SBh, SBh and RP acknowledge support from CEFIPRA project No. 5604-4 and the LIA France-India agreement. HLC and PF acknowledge support from the U.S. Department of Energy, Office of Science, Office of Nuclear Physics under Contract No. DE-AC02-05CH11231 (LBNL). RMPV acknowledge partial support by Ministry of Science, Spain, under the grants BES-2012-061407, SEV-2014-0398, FPA2017-84756-C4 and by EU FEDER funds. AJ was supported by the Spanish Ministerio de Econom\'ia y Competitividad under contract FPA2017-84756-C4-2-P.

 \bibliography{BiblioSb.bib} 
 \bibliographystyle{myapsrev4-1}

\end{document}